\documentclass[aps,prb,reprint,twocolumn,groupedaddress]{revtex4-2}

\usepackage{newtxtext,newtxmath}
\usepackage{graphicx}
\usepackage{bm}
\usepackage{hyperref}
\usepackage{amsmath}
\usepackage[table]{xcolor}
\usepackage{array}
\usepackage{multirow}
\usepackage{physics}

\begin{document}

\title{Quantum melting a Wigner crystal into Hall liquids}

\author{Aidan P. Reddy}
\email{areddy@mit.edu}
\affiliation{Department of Physics, Massachusetts Institute of Technology, Cambridge, Massachusetts 02139, USA}
\author{Liang Fu}
\affiliation{Department of Physics, Massachusetts Institute of Technology, Cambridge, Massachusetts 02139, USA}

\date{\today}

\begin{abstract}

Recent experiments have shown that, counterintuitively, applying a magnetic field to a Wigner crystal can induce quantum Hall effects. In this Letter,
we use variational Monte Carlo to show that a magnetic field can melt zero-field Wigner crystals into integer quantum Hall liquids. This melting originates from quantum oscillations in the liquid's ground state energy, which develops downward cusps at integer filling factors due to incompressibility. Our calculations establish a range of densities in which this quantum melting transition occurs.

\end{abstract}
\maketitle

\emph{Introduction.} When Coulomb interactions overwhelm kinetic energy, an electron gas inevitably crystallizes \cite{wigner1934interaction,monarkha2012two}.
According to quantum Monte Carlo calculations \cite{ceperley1978ground,tanatar1989ground,attaccalite2002correlation,drummond2009quantum,smith2024unified,azadi2024quantum},
this zero-temperature freezing transition occurs at an interaction strength of $r_s \approx 37$ in two dimensions. Evidence for Wigner crystals has been observed in several material systems at both strong \cite{andrei1988observation,stormer1989comment,goldman1990evidence,williams1991conduction,buhmann1991novel,li1991low,paalanen1992electrical,tsui2024direct} and zero  \cite{yoon1999wigner,qiu2012connecting, hossain2020observation,smolenski2021signatures,falson2022competing,munyan2024evidence,xiang2025imaging} magnetic fields.

How does an out-of-plane magnetic field influence electron crystallization?
Applying a magnetic field $B$ to a Wigner crystal monotonically reduces its electrons' quantum zero-point motion
\cite{lozovik1975crystallization,ulinich1979phase}.
In contrast, applying a magnetic field to a Fermi liquid induces quantum Hall states accompanied by quantum oscillations that are periodic in $1/B$.
Because electron liquids and crystals respond differently to magnetic fields, the liquid-crystal phase boundary should depend strongly on field strength, calling for theoretical and experimental study.

\begin{figure}
\centering
\includegraphics[width=\linewidth]{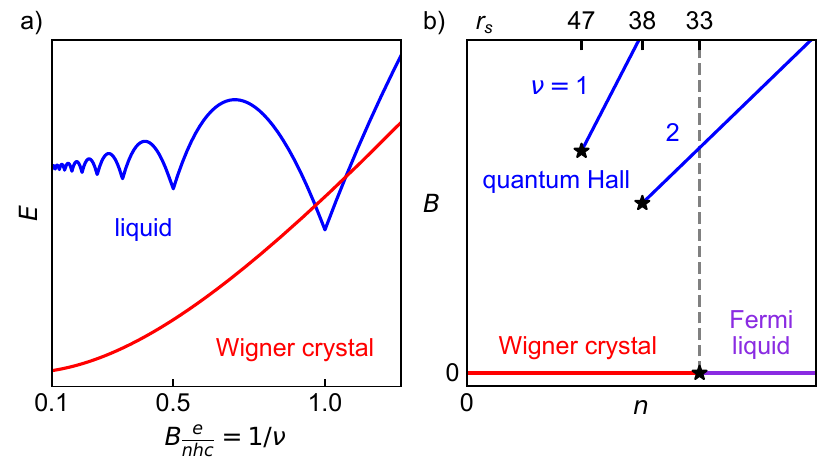}
\caption{\textbf{Quantum melting a Wigner crystal into Hall liquids.} (a) Schematic energies of liquid and crystal phases versus magnetic field $B$, at a density slightly below the zero-field crystallization point. (b) $T=0$ phase diagram of the fully spin-polarized 2DEG in the $n - B$ plane. The stars mark phase boundaries estimated from our variational Monte Carlo at $\nu=1,2$ and $B=0$. The dashed line marks the zero-field crystallization density.}
\label{fig:schematicPhaseDiagram}
\end{figure}

In this Letter, we apply variational Monte Carlo (VMC) to determine the quantum phase diagram of the fully spin-polarized  two-dimensional electron gas (2DEG) at integer Landau level fillings $\nu$. We use a unified Slater-Jastrow wavefunction to describe both electron crystal and Hall liquid phases.
At $\nu=1$ and $2$, we find phase transitions between integer quantum Hall states and electron crystals at critical interaction strengths of $r_s\approx 47$ and $\approx 38$, significantly larger than the value $r_s\approx 33$ for the transition between the fully spin-polarized Fermi liquid and Wigner crystal
at $B=0$ \cite{azadi2024quantum,drummond2009quantum}.
Thus, our study reveals that
quantum Hall states penetrate into the Wigner crystal region of the $n - B$ plane, as shown in Fig. \ref{fig:schematicPhaseDiagram}(b).

Our theory explains the puzzling observation in ZnO-based 2DEGs that applying a magnetic field to a Wigner crystal can induce quantum Hall effects \cite{falson2022competing}.

Additionally, we show that, at long wavelengths, the static structure factor is directly tied to the magnetoplasmon dispersion relation and shows universal behavior across the Hall liquid and Wigner crystal phases.

Before presenting detailed calculations, in Fig. \ref{fig:schematicPhaseDiagram}(a) we sketch the ground state energy $E$ of liquid and crystal phases as a function of $B$ at a density slightly below
the onset of Wigner crystallization at $B=0$. The liquid's energy
oscillates as a function of $\nu\propto 1/B$ with downward cusps at integer $\nu$ due to the formation of incompressible quantum Hall states. In contrast, the crystal's energy
is expected to increase smoothly as a function of $B$ due to an increase in zero-point energy. As a result of the different behaviors in $E(B)$,
the liquid's energy dips below the crystal's in a range of magnetic fields near integer $\nu$, causing a transition from crystal to liquid states. This implies a density window in which a Wigner crystal present at $B=0$ melts into quantum Hall liquids, as illustrated in Fig. \ref{fig:schematicPhaseDiagram}(b).

\emph{Model.} We consider a system of spinless electrons in two dimensions interacting through a Coulomb potential in a homogeneous, out-of-plane magnetic field. This system has the Hamiltonian
\begin{align}\label{eq:Ham}
\begin{split}
H &= \sum_{i}\frac{\bm \pi_i ^2}{2m} + \sum_{i<j}\frac{e^2}{|\bm{r}_i-\bm{r}_j|}
\end{split}
\end{align}
where $\bm \pi_i = \bm{p}_i+\frac{e}{c}\bm{A}(\bm{r}_i)$ and $\bm{\nabla} \times \bm{A}=-B\hat{z}$. Full spin polarization is likely to occur at densities and magnetic field strengths of interest.

Our system involves three independent energy scales:
the kinetic energy $\hbar^2 n/m$ (where $n$ is the electron density), the interaction energy $e^2\sqrt{n}$, and the cyclotron energy $\hbar \omega_c$ with cyclotron frequency $ \omega_c=\frac{eB}{mc}$. The ratio of the interaction and kinetic energies defines the dimensionless interaction strength parameter:
$r_s = 1/\sqrt{\pi a_B^2n}$ where $a_B = \frac{\hbar^2}{e^2m}$ is the Bohr radius.
Similarly, the ratio of interaction and cyclotron energies defines the Landau level mixing parameter
\begin{eqnarray}
\kappa = \frac{e^2}{\ell\hbar\omega_c} = r_s \sqrt{\frac{\nu}{2}}
\end{eqnarray}
where $\ell = \sqrt{\frac{\nu}{2\pi n}}=\sqrt{\frac{\hbar c}{eB}}$ is the magnetic length and $\nu=2\pi\ell^2 n $ is the Landau level filling factor. The two dimensionless parameters $r_s$ and $\nu$ determine our system's phase diagram. We emphasize that, unlike at fractional filling where crystallization can, in principle, occur at infinitesimally weak interaction \cite{maki1983static,lam1984liquid,levesque1984crystallization,zhu1993wigner,ortiz1993new,price1993fractional, zhu1995variational,archer2013competing,zhao2018crystallization}, crystallization at integral $\nu$ requires strong interactions and Landau level mixing.

\begin{figure}
\centering
\includegraphics[width=.9\linewidth]{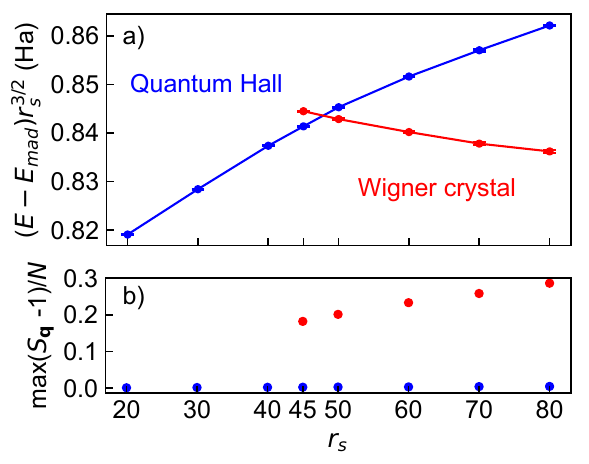}
\caption{\textbf{Quantum phase diagram at $\nu=1$.} (a) Variational energies per particle obtained from the quantum Hall liquid and Wigner crystal \textit{Ansätze} as a function of $r_s$, indicating a phase transition near $r_s\approx 47$. (b) Maximum of the static structure factor of the two \textit{Ansätze}, showing the absence and presence of long-range crystalline order respectively.
$N=144$. Error bars indicate one standard error from binning analysis and $E_{\text{mad}}=-1.106\,103 \text{ Ha}/r_s$ is the Madelung energy.}
\label{fig:nu1PhaseDiagram}
\end{figure}

\emph{Hartree-Fock and Lindemann analysis.} Before presenting our VMC analysis, we summarize two simpler approaches to the problem at hand: Hartree-Fock theory and the Lindemann melting criterion. Hartree-Fock theory predicts a liquid-to-crystal transition at $r_s\approx 2.7$ when $B=0$ \cite{bernu2008metal} and $r_s\approx 9.6$ when $\nu=1$ \cite{SM,smolenski2021signatures}.
However, both of these values are much smaller than the estimates of $r_s\approx 33$ from diffusion Monte Carlo calculations at $B=0$ \cite{azadi2024quantum} and $r_s\approx 47$ from our VMC calculations at $\nu=1$. We present a Hartree-Fock phase diagram for $\nu=1,2,...,7$ in the Supplemental Material, which predicts that the critical $r_s$ increases as $\nu$ decreases. The Lindemann melting criterion asserts that the Wigner crystal melts when $\gamma = \sqrt{\langle \bm{u}^2\rangle}/a_{\text{WC}}$ exceeds a critical value $\gamma^*$, where $a_{\text{WC}}$ is the lattice constant and $\bm u$ is the displacement of an electron's position from its lattice site. This heuristic fails to capture magnetic-field-induced melting because,  within the harmonic approximation, $\sqrt{\langle \bm u^2\rangle}$ monotonically decreases as a function of $B$ \cite{SM,ulinich1979phase}. These theories' shortcomings call for a more accurate approach.

\emph{Variational Monte Carlo methods.}To model both liquid and crystal states, we use a unified Slater-Jastrow wavefunction with a homogeneous two-body Jastrow factor:
\begin{align}
\begin{split}
\Psi(R) &=  \det\left[\psi_i(\bm{r}_j)\right]
\exp\left(\sum_{i<j}u(\bm{r}_i-\bm{r}_j)\right)
\end{split}
\end{align}
where $R=(\bm r_1,...,\bm{r}_N)$ and $N$ is the number of electrons.
We impose the magnetic quasi-periodic boundary condition
$t_i(\bm{L}_{1,2})\Psi(R) = \Psi(R)$ where $\bm L_1$ and $\bm L_2$ are primitive supercell lattice vectors and $t_i$ is a magnetic translation operator for particle $i$ \cite{haldane1985many,brown1964bloch,zak1964magnetic}.
To satisfy this boundary condition, we expand each single-particle orbital $\ket{\psi_{\bm k n}}$ in the Slater determinant as a linear combination of Landau level orbitals $\ket{\phi_{\bm k n}}$, which comprise an orthogonal basis for the boundary condition. Concretely, we write
\begin{align}\label{eq:orbitalRotation}
\begin{split}
\ket{\psi_{\bm{k}m}} = \sum_{n=0}^{n_{\text{max}}} U^{\bm k}_{m n}\ket{\phi_{\bm{k}n}}
\end{split}
\end{align}
where $U^{\bm k}_{m n}$ is a unitary orbital rotation matrix \cite{SM}. Here $n$ is a Landau level index and $\bm{k}$ is a wavevector in the magnetic Brillouin zone.
Each row of the Slater matrix $\psi_i(\bm{r}_j)$ corresponds to one of the orbitals $\ket{\psi_{\bm km}}$, for $0\leq m < \nu$ and every $\bm{k}$. This construction yields the most general Slater determinant, up to the Landau level cutoff $n_{\text{max}}$,  consistent with the magnetic unit cell that we choose to match the Wigner crystal.

The variational parameters of our wavefunction are contained in $u(\bm r)$ and the orbital rotation matrix $U^{\bm{k}}_{mn}$. Using the stochastic reconfiguration algorithm \cite{sorella2001generalized,becca2017quantum}, we optimize these parameters to minimize the energy expectation value of $\Psi(R)$. For each $r_s$, we initialize and constrain the parameters in two ways. For the liquid phase, we optimize only $u(\bm r)$ and set $U^{\bm{k}}_{mn}=\delta_{m n}$ so that the determinant is the non-interacting ground state. For the Wigner crystal phase, we generate crystal ``seed" orbitals by optimizing a randomly initialized determinant with a Jastrow factor at large $r_s$. For other values of $r_s$, we initialize the orbitals to this seed and then optimize them together with the Jastrow factor. In the Supplemental Material, we benchmark our results with exact diagonalization on small system sizes and with a Slater-Jastrow wavefunction of Gaussian orbitals, which we find performs similarly.

\emph{Variational Monte Carlo phase diagram.} In Fig.~\ref{fig:nu1PhaseDiagram}(a), we show variational energies obtained from these two \textit{Ansätze} at $\nu=1$ and several values of $r_s$. When $r_s \geq 50$, the crystal has lower energy, while for $r_s\leq 45$, the liquid is favored. A linear interpolation of the liquid and crystal energies between $r_s=45$ and $50$ locates the melting transition at $r_s\approx 47$.
We show analogous data at $\nu=2$ that indicate a phase transition near $r_s\approx 38$ in the Supplemental Material~\cite{SM}.

Fig. \ref{fig:nu1PhaseDiagram}(b) shows that the variational ground state for $r_s \gtrapprox 50$ has Bragg peaks, indicating crystallization.
Our ansatz assumes magnetic translation symmetry with respect to the Wigner crystal lattice and is therefore incapable of describing microemulsion or other intermediate phases in which translation symmetry is broken at a longer length scale \cite{spivak2003phase,spivak2004phases,sung2025electronic}. Bearing this caveat in mind, the cusp in the ground state energy and sharp discontinuity in the ground state Bragg peak height at the phase boundary are consistent with a first-order melting transition.

The most recent diffusion Monte Carlo calculations available estimate the Fermi-liquid-to-Wigner-crystal transition in the fully spin-polarized 2DEG at $B=0$ to occur at $r_s\approx 33$ \cite{azadi2024quantum} (lower than $r_s \approx 37$ in the presence of spin \cite{smith2024unified,azadi2024quantum}). To obtain a fairer comparison with our VMC results at $B\neq0$, we perform a VMC calculation at $B=0$ and find $r_s\approx 33$ \cite{SM}, in agreement with Ref.~\cite{azadi2024quantum}. This is significantly lower than the values $r_s\approx 47$ and $ 38$ we find at $\nu=1$ and $2$. Therefore, there exists a density window where a Wigner crystal forms at $B=0$ but melts into quantum Hall liquids at small integer fillings.

\begin{figure}
\centering
\includegraphics[width=1.0\linewidth]{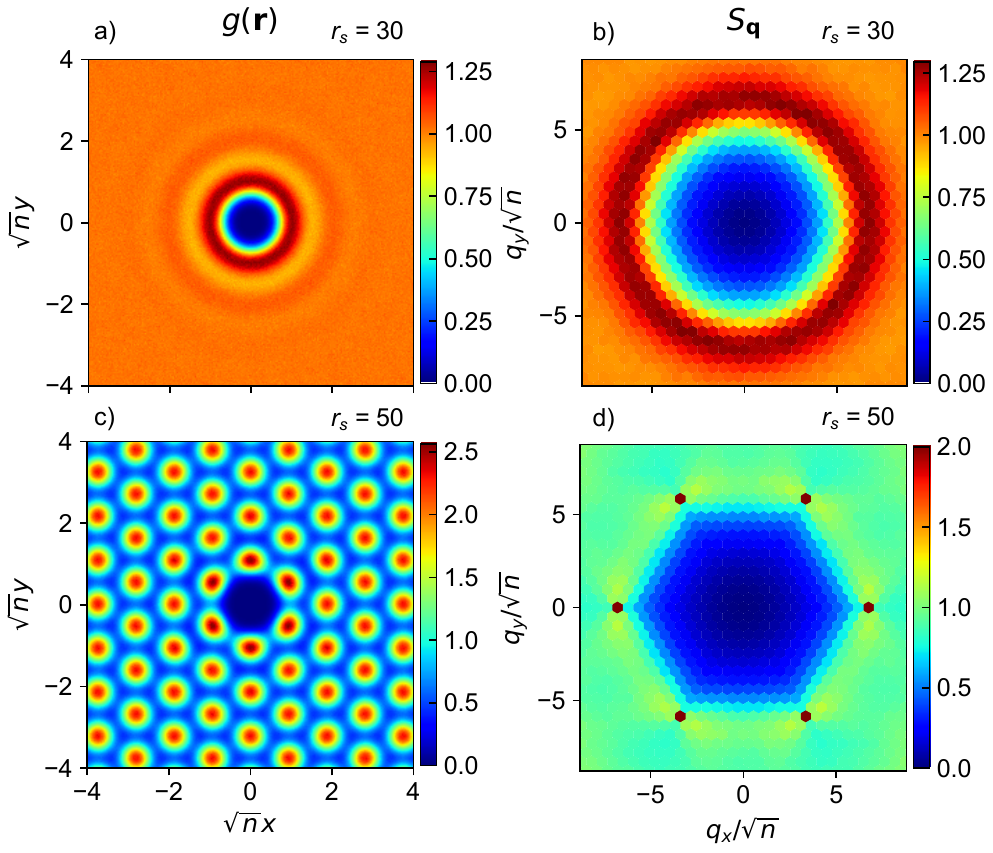}
\caption{\textbf{Density correlations at $\nu=1$.} Pair correlation function (left) and static structure factor (right) of the optimized wavefunctions in the liquid (a,b) and crystal (c,d) phases. The color scale in (d) saturates and $S_{\bm q } \approx 30$ at the first shell of Bragg peaks (dark red points). $N=144$.}
\label{fig:nu1Corr}
\end{figure}

\emph{Density correlations and collective excitations.} Having established our system's phase diagram, we now examine the properties of its liquid and crystal phases. In Fig. \ref{fig:nu1Corr}(a,b) we show density correlation functions of the optimized wavefunction at $r_s=30$. The static structure factor,
\begin{align}
    \begin{split}
        S_{\bm q} = \frac{1}{N}\sum_{i,j}\bra{\Psi}e^{i\bm q \cdot (\bm{r}_i -\bm{r}_j)}\ket{\Psi},
    \end{split}
\end{align}
peaks smoothly around the reciprocal lattice vector  $Q_{\text{WC}} =\sqrt{\frac{8\pi^2}{\sqrt{3}}n}$  of a Wigner crystal. Similarly, the pair correlation function,
\begin{align}
    \begin{split}
        g(\bm r) &= \frac{A}{N^2} \sum_{i\neq j}\bra{\Psi}\delta^2(\bm r-\bm r_i + \bm{r}_j)\ket{\Psi},
    \end{split}
\end{align}
peaks smoothly around the lattice constant $a_{\text{WC}} = \sqrt{\frac{2}{\sqrt{3}n}}$ of a Wigner crystal. These features are typical of a strongly correlated liquid. In contrast, the density correlation functions at $r_s=60$ shown in Fig. \ref{fig:nu1Corr}(c,d) reveal crystallization. The real-space pair correlation function exhibits periodic modulation throughout the system. This long-range order appears as Bragg peaks in the static structure factor at the reciprocal lattice vector $Q_{\text{WC}}\hat{x}$ and its $\pi/3$ rotations.

We now relate our system's long-wavelength density-density correlations to its collective excitations. Classically, a two-dimensional electron liquid in an out-of-plane magnetic field interacting through a Coulomb potential $v(q) = 2\pi e^2/q$ has a longitudinal collective excitation mode, the magnetoplasmon, with a dispersion relation
\begin{equation}
\begin{split}
\omega_{p}(q) &=\sqrt{\omega_c^2 + \frac{2\pi e^2 n}{m}q}.
\end{split}
\end{equation}
Quantum mechanically, the magnetoexciton of the integer quantum Hall state and longitudinal phonon of the Wigner crystal follow this dispersion relation at small $q$, as can be shown rigorously in the limits of small and large $r_s$ \cite{kallin1984excitations,macdonald1985hartree,bonsall1977some}. In the limit $B\rightarrow 0$, the plasmon dispersion of a 2D Coulomb gas, $\omega_p(q)= \sqrt{\frac{2\pi e^2 n}{m}q}$, is recovered.

Let us assume that this collective mode exists in the quantum system and exhausts the $f$-sum rule at generic $r_s$. Under this assumption, the static structure factor obeys
\begin{align}
\begin{split}
S_{q } &= \frac{\hbar q^2}{2m\omega_{p}(q)} = \frac{\ell^2}{2} q^2 -\frac{\kappa\nu\ell^3}{4} q^3+ \mathcal{O}(q^4)
\end{split}
\end{align}
at small $q$. The leading $\mathcal{O}(q^2)$ term is an exact property of the quantum system guaranteed by Kohn's theorem
\cite{kohn1961cyclotron} -- that is, the optical spectral weight
is completely exhausted by the cyclotron mode due to the decoupling between electrons' center-of-mass and relative degrees of freedom.
In the quantum Hall phase, the property $\lim_{q\rightarrow 0} S_q/q^2 \geq \frac{\ell^2}{2}$ is also required by the system's topologically quantized Hall conductance $\sigma_H=\nu e^2/h$ \cite{onishi2024topological,onishi2025quantum}. The subleading $\mathcal{O}(q^3)$ term, which is non-analytic as a function of $\bm{q}$, is a consequence of the Coulomb interaction in gapped two-dimensional systems.

\begin{figure}
\centering
\includegraphics[width=1.\linewidth]{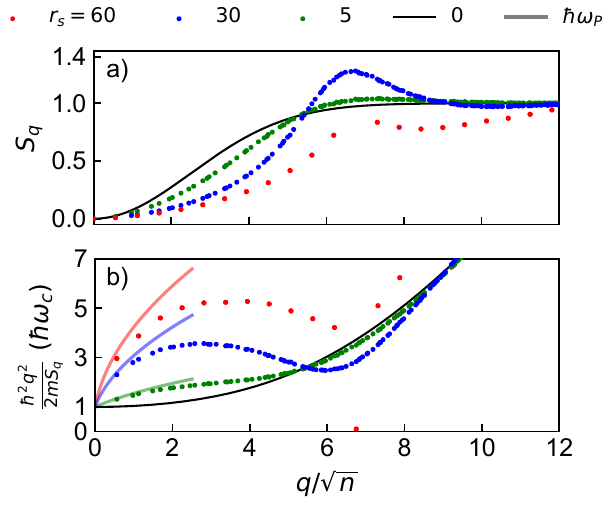}
\caption{\textbf{Static structure factor and magnetoplasmons.} (a) Static structure factor $S_q$ of the variational ground state at $\nu=1$. (b) Dispersion of longitudinal collective modes from the single-mode approximation, which converges with the the classical magnetoplasmon dispersion (semi-transparent lines) at small $q$. $r_s=0$ curves are analytical. The data are rotation-averaged in the liquid phase and taken along the $x$-axis in the crystal phase. $N=144$.}
\label{fig:magnetoplasmon}
\end{figure}

In Fig. \ref{fig:magnetoplasmon}, we test this assumption by comparing $\hbar \omega_p(q)$ with the single-mode approximation dispersion, $\Delta(q) \equiv \frac{\hbar^2 q^2}{2m S_q}$, across a wide range of interaction strengths \cite{feynman1954atomic}. $\hbar\omega_p(q)$ and $\Delta(q)$ converge as $q \rightarrow 0$ in both the liquid and Wigner crystal phases. This supports the existence of a magnetoplasmon mode that exhausts the $f$-sum rule in the limit $q\rightarrow 0$ for all $r_s$. It also shows that our wavefunctions respect Kohn's theorem. We emphasize that this behavior is not enforced by construction but rather obtained through optimization, signaling that our wavefunctions are accurate. Additionally, when $r_s$ is large, $\Delta(q)$ shows a roton minimum near $Q_{\text{WC}}$. This rather generic feature of strongly interacting quantum liquids is a direct consequence of the peak in $S_q$ and reflects strong short-range correlations \cite{feynman1954atomic,girvin1986magneto}.

\emph{Discussion.} We have shown that a magnetic field can melt a Wigner crystal into quantum Hall liquids at small integer Landau level fillings. Using the variational Monte Carlo method, we have computed the quantum phase diagram of the fully spin-polarized 2DEG at $\nu=1$ and $2$. We find phase transitions between integer quantum Hall and Wigner crystal states at $r_s\approx 47$ and $\approx 38$, respectively. These interaction strengths are larger than the $B=0$ transition points, $r_s\approx 37$ with spin and $r_s\approx 33$ without spin \cite{smith2024unified,azadi2024quantum}.
This establishes a density window in which a Wigner crystal present at $B=0$ melts into a quantum Hall state under a magnetic field. Experimentally, this implies that at fixed density the system
is strongly insulating at zero field but exhibits a quantized Hall effect with $R_{xx}=0$ at small integer $\nu$, as observed in Ref. \cite{falson2022competing}.
Finally, we have shown that magnetoplasmon collective excitations control the system's long-wavelength density correlations, which exhibit unified behavior across both liquid and crystal phases.

While we have focused on integer $\nu$, similar magnetic-field-induced transitions between liquid and crystal phases also occur at fractional Landau level filling. For example, at small $\nu<1$, electron crystal phases are interrupted by fractional quantum Hall liquids at special fractions such as $\frac{1}{3}$ or $\frac{1}{5}$, as has been observed in multiple material systems \cite{jiang1990quantum,goldman1990evidence,li1991low,jiang1991magnetotransport,santos1992observation,maryenko2018composite,chen2019competing,ma2020thermal,villegas2021competition,tsui2024direct,dong2025nonlinear,haug2025interaction} and studied theoretically \cite{zhu1995variational,archer2013competing,zhao2018crystallization}. When these liquid-crystal oscillations occur in higher Landau levels, they lead to ``reentrant integer quantum Hall effects'' \cite{eisenstein2002insulating,xia2004electron,kumar2010nonconventional,liu2012observation,deng2012collective,zhou2020solids}. In some moiré systems at $B=0$, integer quantum anomalous Hall states are observed over a broad range of densities,
except certain fractional band fillings where fractional quantum anomalous Hall states form \cite{lu2025extended,xu2025signatures}. These phenomena likely also originate from downward cusps in the liquid's energy due to incompressibility that bring it below the energy of competing crystal phases (see Fig. \ref{fig:schematicPhaseDiagram}(b)).

The possible coexistence of quantized Hall effect with crystalline order in a \emph{Hall crystal} phase has been proposed theoretically \cite{tevsanovic1989hall}. However, we find no evidence for such a state at $\nu=1$ given a standard Coulomb interaction, in agreement with the expectations of Ref. \cite{tevsanovic1989hall}.

Our work invites further research in related directions.
One natural extension is to explore the influence of a moiré potential, which can give rise to generalized Wigner crystals \cite{fedorovich2025first,zong2025quantum,chen2023magnetic,morales2023magnetism,zhou2024quantum,li2021imaging,regan2020mott,xu2020correlated}. Another compelling direction is to investigate
interplay between magnetic fields and Wigner crystallization in few-layer graphene systems, which feature unconventional band dispersions and large Berry curvature \cite{tsui2024direct}. Crystallization at $B=0$ in flat bands with Berry curvature is also an intriguing possibility to address numerically beyond the Hartree-Fock approximation \cite{joy2023wigner,dong2024theory,zhou2024fractional,dong2024anomalous,tan2024parent,dong2024stability,tan2025variational,soejima2024anomalous,reddy2023fractional,reddy2023toward,sheng2024quantum,joy2025chiral,kim2025magnetic}. More broadly, microscopic mechanisms for coexistence between crystallization and topology—whether at zero or finite magnetic field—remain to be explored \cite{kivelson1986cooperative,tevsanovic1989hall}. Finally, it may be possible to combine aspects of our wavefunction ansatz with neural networks to study competing liquid and crystal phases in quantum Hall systems and Chern bands \cite{teng2025solving,qian2025describing, li2025deep,luo2025solving}.

\emph{Acknowledgments.} We thank Max Geier, Yaar Vituri, Bryan Clark, Shiwei Zhang, Agnes Valenti, David Dai, Patrick Ledwith, and Khachatur Nazaryan for helpful discussions. AR is particularly grateful to Ahmed Abouelkomsan for numerous helpful discussions and encouragement.
The authors acknowledge the MIT SuperCloud and Lincoln Laboratory Supercomputing Center for providing computing resources that have contributed to the research results reported in this paper.
This work was supported by the U.S. Army DEVCOM ARL Army Research Office through the MIT Institute for Soldier Nanotechnologies under Cooperative Agreement number W911NF-23-2-0121. LF was supported by a Simons Investigator Award from the Simons Foundation.

\emph{Data availability.} The data and code that support the ﬁndings of this article are openly available \cite{reddy_zenodo}.

\end{document}


\title{Supplemental Material for \emph{Quantum melting a Wigner crystal into Hall liquids}}

\author{Aidan P. Reddy}
\email{areddy@mit.edu}
\affiliation{Department of Physics, Massachusetts Institute of Technology, Cambridge, Massachusetts 02139, USA}

\author{Liang Fu}
\affiliation{Department of Physics, Massachusetts Institute of Technology, Cambridge, Massachusetts 02139, USA}

\date{\today}

\maketitle
\tableofcontents

\section{Extended results}

\subsection{Phase diagrams at $\nu=2$ and $B=0$}

\begin{figure}
    \centering
    \includegraphics[width=.8\linewidth]{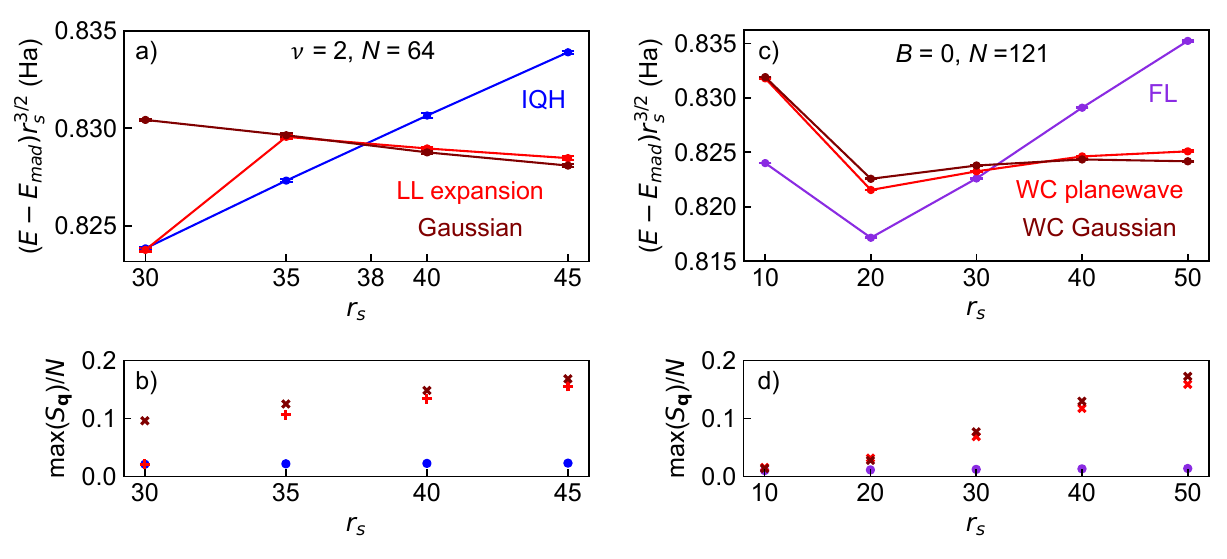}
    \caption{\textbf{Phase diagrams at $\nu=2$ and $B=0$.} Variational energies (a,c) and maxima of the static structure factor $S_{\bm q}$ (b,d) at $\nu=2$ (a,b) and $B=0$ (c,d).
    At $\nu=2,\, r_s=30$ the Landau level expansion wavefunction flows through optimization to a liquid wavefunction that is essentially identical to the one obtained from liquid initialization, as is evident from their matching $\text{max}(S_{\bm q})$ values and energies (within error bars). In contrast, at $B=0$, we use different $k$-space meshes (see Sec. \ref{subsec:planewave}) for the liquid and crystal ansatz, so the crystal ansatz cannot flow to a liquid state through optimization. $N=64$ in (a,b) and $N=121$ in (c,d). As always, we assume full spin polarization. FL = Fermi liquid, WC = Wigner crystal.}
    \label{fig:nu2B0PhaseDiagram}
\end{figure}

In Fig. \ref{fig:nu2B0PhaseDiagram}, we show variational energies at $\nu=2$ and at $B=0$. The data at $\nu=2$ shown in Fig. \ref{fig:nu2B0PhaseDiagram}(a) indicate a quantum Hall to Wigner crystal transition near $r_s\approx 38$. The data at $B=0$ in Fig. \ref{fig:nu2B0PhaseDiagram}(c) indicate a Fermi liquid to Wigner crystal transition near $r_s\approx 33$. We have compared energies from the integer quantum Hall and Landau level basis expansion (or, at $B=0$, plane wave expansion) ans\"atze described in the main text with Gaussian orbitals, which we explain in Sec. \ref{subsec:gaussian}. We find that the variational wavefunctions with Gaussian orbitals have energies close to those with Landau level or plane wave basis orbitals.

Our estimate of the critical $r_s$ at $B=0$ is in agreement with the value $r_s\approx 33$ implied by Fig. 6 of Ref. \cite{azadi2024quantum}, but is somewhat higher than $r_s\approx 27$ implied by Fig. 1 of the earlier work Ref. \cite{drummond2009quantum}. Both Refs. \cite{azadi2024quantum,drummond2009quantum} used diffusion Monte Carlo. However, Ref. \cite{azadi2024quantum} obtained a higher estimate of the critical $r_s$ than Ref. \cite{drummond2009quantum} because Ref. \cite{azadi2024quantum} found a lower energy for the fully spin-polarized Fermi liquid phase than Ref. \cite{drummond2009quantum} and presented this alongside data for the fully spin-polarized Wigner crystal taken from Ref. \cite{drummond2009quantum}.

The purpose of our $B=0$ calculation is merely to substantiate our conclusion that the liquid-crystal transition tends toward higher $r_s$ as $B$ increases (restricting to integral $\nu$), not to obtain state-of-the-art accuracy. Thus, for simplicity, we do not twist-average, so finite size effects at $B=0$ may be significant \cite{lin2001twist}. As we explain in Section \ref{subsec:magOrbitals}, twist-averaging is entirely unnecessary when $B\neq 0$.

\subsection{Pair correlation function}

\begin{figure}
    \centering
    \includegraphics[width=0.3\linewidth]{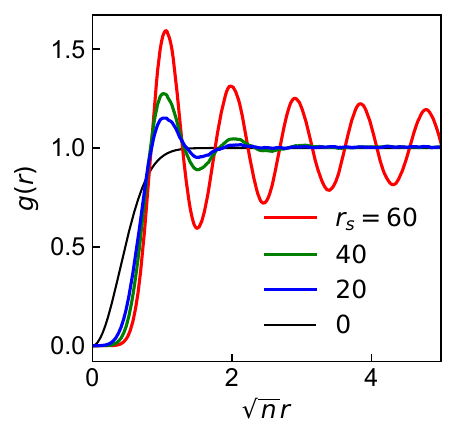}
    \caption{\textbf{Pair correlation function.} Rotation-averaged pair correlation function of the $\nu=1$ ground states at several interaction strengths. The curve at $r_s=0$ is computed analytically and the curves at $r_s > 0$ are taken from sampling optimized wavefunctions (integer quantum Hall ansatz at $r_s=20,40$ and Landau level basis expansion ansatz at $r_s=60$). $N=144$.}
    \label{fig:pairCorr}
\end{figure}

In Fig. \ref{fig:pairCorr}, we show rotation-averaged pair correlation functions of the ground state at $\nu=1$ for several $r_s$ values.

\subsection{Exact diagonalization benchmark for small system size}

\begin{figure}
    \centering
    \includegraphics[width=0.6\linewidth]{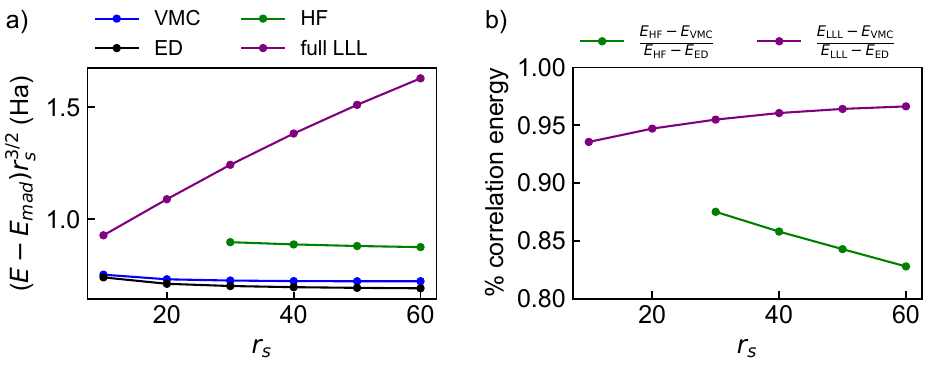}
    \caption{\textbf{Benchmark with exact diagonalization for 4 electrons.} (a) Comparison of energies from self-consistent Hartree-Fock (HF), the full lowest Landau level Slater determinant, exact diagonalization (ED), variational Monte Carlo (VMC) wavefunctions. The exact diagonalization calculation includes 15 Landau levels. We believe that the exact diagonalization energy is converged with respect to Landau level truncation because the weight of the $n=10$ through 14 Landau levels all together in the variational ground state is less than $5\times 10^{-4}$ at $r_s=60$. (b) Percentage of the correlation energy recovered by the VMC wavefunction. We consider two definitions of the correlation energy: one where the ``Hartree-Fock'' energy is defined with respect to the full lowest Landau level Slater determinant and one where it is defined with respect to the self-consistent ground state. In both cases, the percentage of the correlation energy recovered by the VMC wavefunction is high, demonstrating its accuracy.}
    \label{fig:Ne4}
\end{figure}

In Fig. \ref{fig:Ne4}, we compare energies from VMC, Hartree-Fock, and exact diagonalization on a small system with $N=4$ electrons. We show the energies of both the full lowest Landau level Slater determinant and self-consistent Hartree-Fock ground state, which is a crystal for the $r_s$ values shown. For the VMC wavefunction, we find that the integer quantum Hall orbitals are optimal for all $r_s$ studied. This is sensible given that the system is extremely small, blurring the boundary between short and long-range order. Fig. \ref{fig:Ne4} (b) shows that the VMC wavefunction recovers the vast majority of the correlation energy.

\begin{figure}
    \centering
    \includegraphics[width=\linewidth]{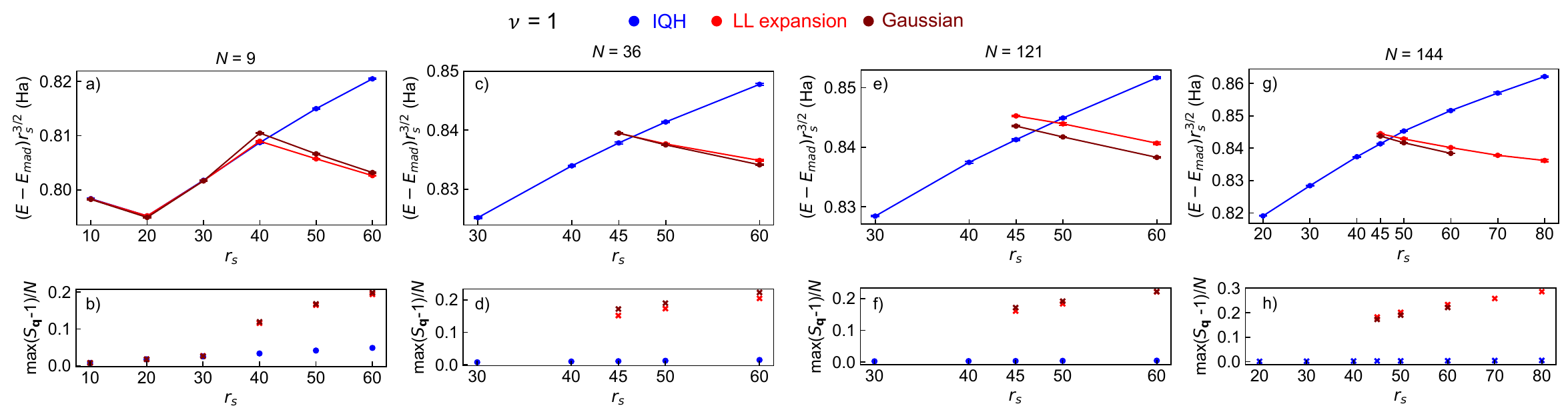}
    \caption{\textbf{Comparison of $\nu=1$ variational energies for multiple ansätze and system sizes.} Energy per particle for various system sizes (top row) and magnitude of structure factor peaks. The values of $r_s$ for the liquid transition we estimate by linearly interpolating the energies for $N=9,36,121$ and $144$ are $\approx 40, 46.5, 47$, and $47$. IQH = integer quantum Hall.}
    \label{fig:nu1comparison}
\end{figure}

In Fig. \ref{fig:nu1comparison}, we show variational energies for systems with $N=9,36,121$ and $144$ electrons for integer quantum Hall, Landau level basis expansion, and Gaussian orbitals. Generally, we find that the Gaussian orbitals perform similarly to the LL basis expansion orbitals. On the larger system sizes $N=121$ and $144$, the Gaussian orbitals outperform the Landau level basis orbitals slightly. However, the consistency in the estimates of the critical $r_s$ across $N=36$, $N=121$, and $N=144$ supports their accuracy.

\newpage

\subsection{Table of variational energies}

\begin{longtable}{|c|c|c|c|l|c|}
\caption{\textbf{Table of energies of optimized variational wavefunctions.} $N$ is the number of electrons. All energies are per particle. $E_{\text{mad}}=-1.106\,103 \text{ Ha}/r_s$ is the Madelung energy. Standard error is determined through binning analysis of local energy Monte Carlo data.}\label{table:energies}\\
\hline
$B$ field & Wavefunction type & $N$ & $\;\; r_s \;\;$ & $(E-E_{\text{mad}}) r_s^{3/2}$ (Ha) & std. err. \\
\hline
\endhead

\hline
\endfoot

\hline
\endlastfoot

\multirow{48}{*}{$\nu=1$} & \multirow{24}{*}{Integer quantum Hall} & \multirow{6}{*}{9} & 10 & 0.79841 & 6e-5 \\
 &  &  & 20 & 0.7952 & 1.1e-4 \\
 &  &  & 30 & 0.80179 & 9e-5 \\
 &  &  & 40 & 0.80874 & 9e-5 \\
 &  &  & 50 & 0.8150 & 1.4e-4 \\
 &  &  & 60 & 0.8205 & 1.4e-4 \\
\cline{3-6}
 &  & \multirow{5}{*}{36} & 30 & 0.8252 & 1.2e-4 \\
 &  &  & 40 & 0.8340 & 1.3e-4 \\
 &  &  & 45 & 0.8378 & 1.9e-4 \\
 &  &  & 50 & 0.8414 & 1.4e-4 \\
 &  &  & 60 & 0.8478 & 1.5e-4 \\
\cline{3-6}
 &  & \multirow{5}{*}{121} & 30 & 0.8284 & 1.3e-4 \\
 &  &  & 40 & 0.8375 & 1.2e-4 \\
 &  &  & 45 & 0.8413 & 1.1e-4 \\
 &  &  & 50 & 0.8449 & 1.4e-4 \\
 &  &  & 60 & 0.8517 & 1.3e-4 \\
\cline{3-6}
 &  & \multirow{8}{*}{144} & 20 & 0.81909 & 7e-5 \\
 &  &  & 30 & 0.8284 & 1.2e-4 \\
 &  &  & 40 & 0.83736 & 9e-5 \\
 &  &  & 45 & 0.8414 & 1.4e-4 \\
 &  &  & 50 & 0.8453 & 1.3e-4 \\
 &  &  & 60 & 0.8516 & 1e-4 \\
 &  &  & 70 & 0.8570 & 2.6e-4 \\
 &  &  & 80 & 0.8621 & 1.9e-4 \\
\cline{2-6}
 & \multirow{14}{*}{Landau level expansion} & \multirow{6}{*}{9} & 10 & 0.79834 & 6e-5 \\
 &  &  & 20 & 0.79524 & 8e-5 \\
 &  &  & 30 & 0.8017 & 1.3e-4 \\
 &  &  & 40 & 0.8090 & 1.1e-4 \\
 &  &  & 50 & 0.80574 & 7e-5 \\
 &  &  & 60 & 0.80266 & 8e-5 \\
\cline{3-6}
 &  & \multirow{3}{*}{36} & 45 & 0.8395 & 1e-4 \\
 &  &  & 50 & 0.8377 & 1e-4 \\
 &  &  & 60 & 0.8349 & 1.4e-4 \\
\cline{3-6}
 &  & \multirow{3}{*}{121} & 45 & 0.8453 & 1.3e-4 \\
 &  &  & 50 & 0.8439 & 2.1e-4 \\
 &  &  & 60 & 0.8407 & 1.8e-4 \\
\cline{3-6}
 &  & \multirow{5}{*}{144} & 45 & 0.84448 & 9e-5 \\
 &  &  & 50 & 0.8429 & 1.9e-4 \\
 &  &  & 60 & 0.8402 & 1.7e-4 \\
 &  &  & 70 & 0.8378 & 2.9e-4 \\
 &  &  & 80 & 0.8362 & 4e-4 \\
\cline{2-6}
 & \multirow{10}{*}{Gaussian} & \multirow{6}{*}{9} & 10 & 0.79828 & 9e-5 \\
 &  &  & 20 & 0.7950 & 1.2e-4 \\
 &  &  & 30 & 0.8017 & 1.1e-4 \\
 &  &  & 40 & 0.81048 & 8e-5 \\
 &  &  & 50 & 0.8067 & 1.1e-4 \\
 &  &  & 60 & 0.80322 & 8e-5 \\
\cline{3-6}
 &  & \multirow{3}{*}{36} & 45 & 0.8395 & 1e-4 \\
 &  &  & 50 & 0.83751 & 8e-5 \\
 &  &  & 60 & 0.83413 & 9e-5 \\
\cline{3-6}
 &  & \multirow{3}{*}{121} & 45 & 0.84355 & 7e-5 \\
 &  &  & 50 & 0.84173 & 5e-5 \\
 &  &  & 60 & 0.83830 & 8e-5 \\
\cline{3-6}
 &  & \multirow{3}{*}{144} & 45 & 0.84373 & 9e-5 \\
 &  &  & 50 & 0.8417 & 1.1e-4 \\
 &  &  & 60 & 0.83841 & 6e-5 \\
\hline
\multirow{12}{*}{$\nu=2$} & \multirow{4}{*}{Integer quantum Hall} & \multirow{4}{*}{64} & 30 & 0.82384 & 7e-5 \\
 &  &  & 35 & 0.82731 & 9e-5 \\
 &  &  & 40 & 0.8307 & 1.2e-4 \\
 &  &  & 45 & 0.83389 & 8e-5 \\
\cline{2-6}
 & \multirow{4}{*}{Landau level expansion} & \multirow{4}{*}{64} & 30 & 0.82376 & 8e-5 \\
 &  &  & 35 & 0.8296 & 1.2e-4 \\
 &  &  & 40 & 0.82896 & 9e-5 \\
 &  &  & 45 & 0.8285 & 1e-4 \\
\cline{2-6}
 & \multirow{4}{*}{Gaussian} & \multirow{4}{*}{64} & 30 & 0.83043 & 6e-5 \\
 &  &  & 35 & 0.8296 & 1e-4 \\
 &  &  & 40 & 0.82876 & 8e-5 \\
 &  &  & 45 & 0.82808 & 9e-5 \\
\hline
\multirow{15}{*}{$B=0$} & \multirow{5}{*}{Fermi liquid} & \multirow{5}{*}{121} & 10 & 0.82401 & 2.2e-5 \\
 &  &  & 20 & 0.81714 & 3e-5 \\
 &  &  & 30 & 0.82260 & 4e-5 \\
 &  &  & 40 & 0.82910 & 5e-5 \\
 &  &  & 50 & 0.83524 & 7e-5 \\
\cline{2-6}
 & \multirow{5}{*}{Planewave expansion} & \multirow{5}{*}{121} & 10 & 0.83179 & 2.1e-5 \\
 &  &  & 20 & 0.82154 & 3e-5 \\
 &  &  & 30 & 0.82325 & 6e-5 \\
 &  &  & 40 & 0.82462 & 6e-5 \\
 &  &  & 50 & 0.8251 & 1e-4 \\
\cline{2-6}
 & \multirow{5}{*}{Gaussian} & \multirow{5}{*}{121} & 10 & 0.83191 & 2.8e-5 \\
 &  &  & 20 & 0.82256 & 4e-5 \\
 &  &  & 30 & 0.82378 & 4e-5 \\
 &  &  & 40 & 0.82434 & 5e-5 \\
 &  &  & 50 & 0.82417 & 5e-5 \\
\end{longtable}

\section{Hartree-Fock and Lindemann criterion analysis}

\subsection{Hartree-Fock phase diagram and VMC energy comparison}

Here we study the quantum phase diagram of the spin-polarized 2DEG at integral $\nu$ within the Hartree-Fock approximation. We adopt the same boundary conditions and supercell geometry as in our variational Monte Carlo calculations (see Sec. \ref{subsec:supercell}). We include 14 Landau levels and use supercells containing $\geq 200$ flux quanta for every $\nu$. We compute Landau level basis Coulomb matrix elements following the Supplementary Material of Ref.~\cite{reddy2024non}.

We compare the variational energies obtained by self-consistent iteration starting from three different density matrices: the non-interacting ground state density matrix, a random initial density matrix, and a Wigner crystal ``seed'' density matrix. In all cases, the self-consistent solutions obtained from random initial density matrices have energy higher than or equal to the energies obtained from the other initial density matrices. In the presence of interactions, the non-interacting integer quantum Hall ground state of $\nu$ full Landau levels is not a perfectly self-consistent solution at finite system size. However, we find numerically that it is self-consistent to an excellent approximation at the system sizes used here and we believe it is perfectly self-consistent in the thermodynamic limit.

\begin{figure}
    \centering
    \includegraphics[width=0.5\linewidth]{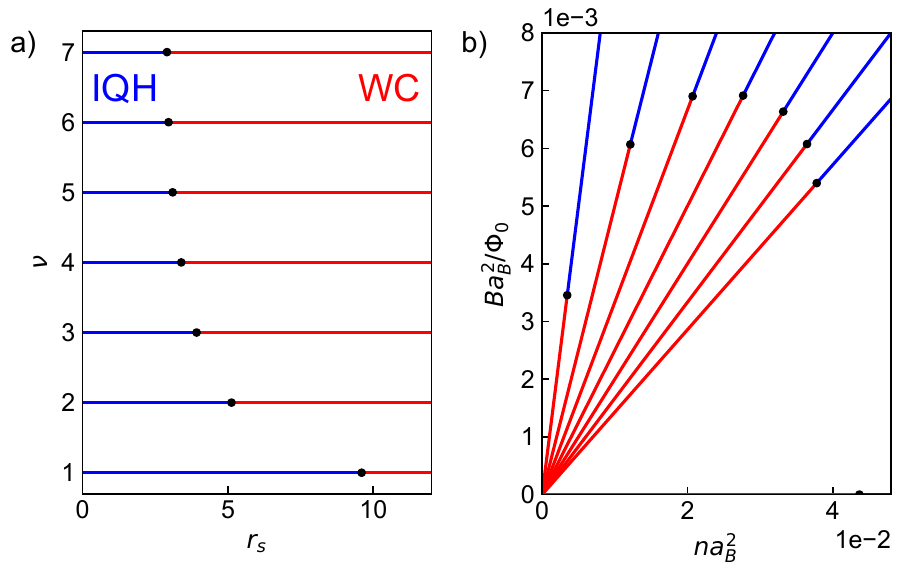}
    \caption{Hartree-Fock quantum phase diagrams of the fully spin-polarized 2DEG at several integral Landau level filling factors $\nu$ in the (a) $\nu - r_s$ plane and (b) $n - B$ planes. For each $\nu$, there are two phases, integer quantum Hall (blue) and Wigner crystal (red), separated by a first-order phase transition (black dots).}
    \label{fig:HFPhaseDiagram}
\end{figure}

In Fig. \ref{fig:HFPhaseDiagram}, we show the Hartree-Fock quantum phase diagram. For each $\nu$, we perform calculations on a regular grid of $r_s$ values with spacing $0.5$.
We estimate the transition points $r^*_{s,\nu}$ (black dots) by making a linear approximation to the energies of the liquid and crystal phases according to their values at the two $r_s$ points closest to the transition. As shown in Fig. ~\ref{fig:HFPhaseDiagram}(a), $r^*_{s,\nu}$ monotonically decreases as $\nu$ increases, approaching the $B=0$ limit of $\approx 2.7$ as $\nu$ increases \cite{bernu2008metal}.

\begin{figure}
    \centering
    \includegraphics[width=0.5\linewidth]{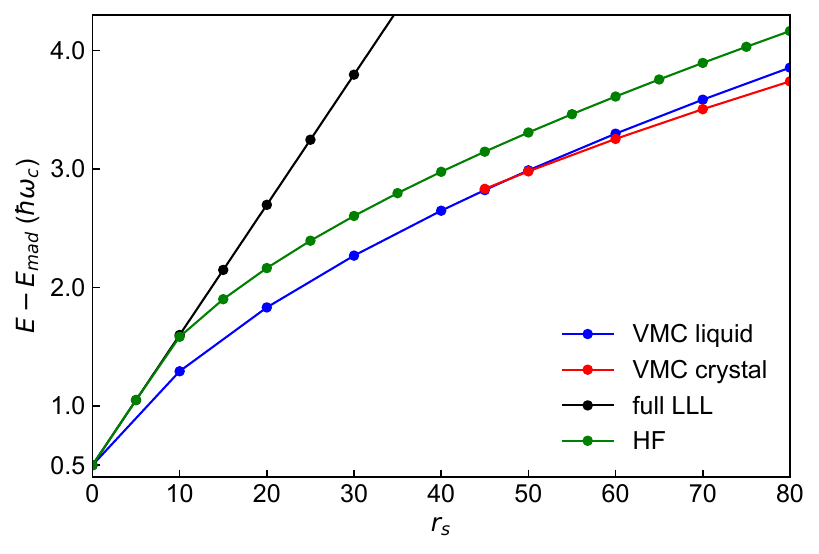}
    \caption{Comparison of Hartree-Fock and variational Monte Carlo energies at $\nu=1$. The ``full LLL'' energy is that of the full lowest Landau Level Slater determinant (the non-interacting ground state). The self-consistent HF energy is the energy of the lowest self-consistent HF state, which is a crystal for $r_s\gtrapprox 9.6$ (see Fig.~\ref{fig:HFPhaseDiagram}). VMC data are the same as in Fig. 1 of the main text and we omit statistical error bars here. $N=144$.}
    \label{fig:HFVMCComparison}
\end{figure}

In Fig.~\ref{fig:HFVMCComparison}, we compare Hartree-Fock and variational Monte Carlo energies at $\nu=1$. Hartree-Fock is strongly biased in favor of the Wigner crystal. The large difference between the energies of the full lowest Landau Level Slater determinant (black) and the variational Monte Carlo liquid state (blue) energies demonstrates significant correlations in the liquid phase.

\subsection{Liquid energy in the Hartree-Fock approximation}

To produce the liquid energy curve in Fig. 1(a) of the main text, we compute the energy expectation value of the thermal density matrix $\rho$ of a non-interacting system in the limit $T\rightarrow 0$. In this density matrix, all degenerate many-body ground states of the kinetic energy operator are equally occupied and all higher energy states are unoccupied. The kinetic energy per particle of this density matrix is
\begin{align}
    \begin{split}
        K &= \frac{\hbar\omega_c}{\nu}\left[\frac{\lfloor\nu\rfloor^2 }{2} + (\nu-\lfloor \nu \rfloor)\left(\lfloor \nu \rfloor + \frac{1}{2}\right)\right].
    \end{split}
\end{align}
The exchange energy per particle is
\begin{align}
    \begin{split}
        \Sigma_X &= -\frac{e^2}{\ell}\frac{1}{2\nu}\sum_{m,n}\nu_{n}\nu_{m} X_{nm}
    \end{split}
\end{align}
where
\begin{align}
    \begin{split}
        X_{nm} &= \frac{\min(n,m)!}{\max(n,m)!}\int_0^{\infty} dx\, e^{-x^2/2}\left(\frac{x^2}{2}\right)^{|m-n|}\left[ L_{\min(n,m)}^{|m-n|} \left(\frac{x^2}{2}\right)\right]^2
    \end{split}
\end{align}
and $\nu_n$ is the occupation of the $n^{th}$ Landau level ($\nu_{n}=1$ if $n < \lfloor \nu \rfloor$ and $=\max(0,\nu-n)$ otherwise). $L_{n}^{m}$ is a generalized Laguerre polynomial. We derived this expression following the discussion of Landau level basis Coulomb matrix elements in the Supplementary Material of Ref.~\cite{reddy2024non}.

At integral $\nu$, $K$ takes the same value for all $\nu$ (in $B$-independent units, such as $E_F \equiv \frac{2\pi \hbar^2 n}{m} = \nu \hbar\omega_c /2$) while $\Sigma_{X}$ tends downward as $\nu$ decreases. Needless to say, this infinitesimal-temperature Hartree-Fock approximation neglects many features of the liquid energy, particularly at non-integral $\nu$. However, it correctly captures a robust feature of the liquid energy that we emphasize in the introduction of the main text: quantum oscillations with cusp minima at integral filling factors.

\subsection{Phonons and Lindemann melting curve}

\begin{figure}
    \centering
    \includegraphics[width=0.8\linewidth]{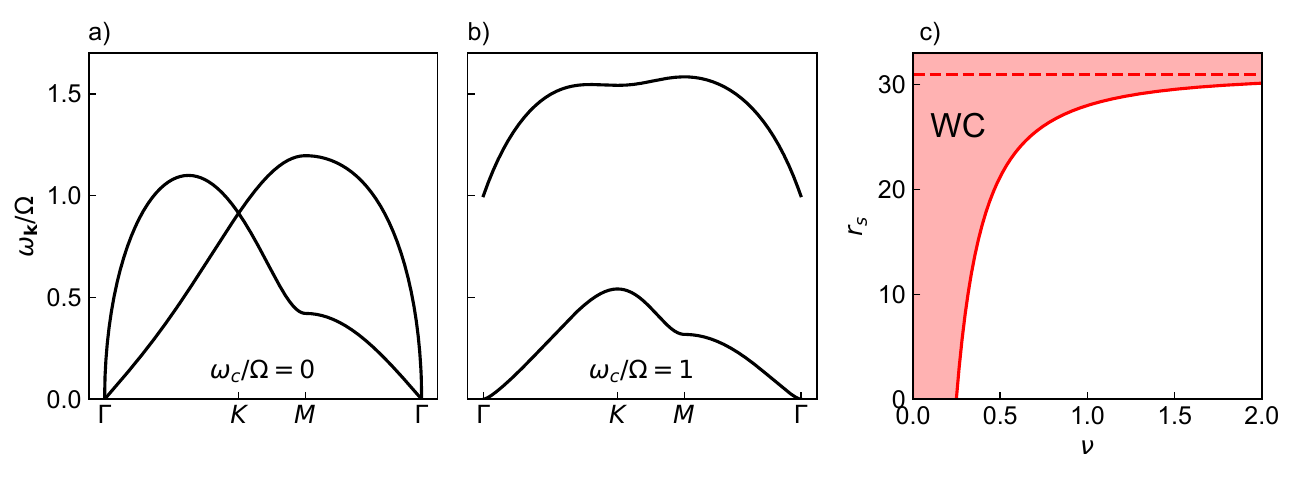}
    \caption{\textbf{Phonons and Lindemann melting criterion.} (a,b) Phonon spectra in the harmonic approximation at (a) $\omega_c=0$ and (b) $\omega_c=\Omega \equiv \sqrt{\frac{8e^2}{ma_{\text{WC}}^3}}$. (c) Melting curve of the Wigner crystal in a magnetic field according to the Lindemann criterion with Lindemann ratio $\gamma^*=0.3$. The pink region is the crystal phase, the solid red line is the melting curve, and the red dashed line marks the melting point at $B=0$ (equivalently, $\nu=\infty$).}
    \label{fig:lindemannCurve}
\end{figure}

Here we apply the heuristic quantum Lindemann criterion to study the boundary of the Wigner crystal phase in a magnetic field. The Lindemann criterion asserts that when the Lindemann ratio $\gamma = \sqrt{\langle \bm{u}^2\rangle}/a_{\text{WC}}$ exceeds a critical value $\gamma^*$ where $\bm u$ is the deviation of an electron from its equilibrium position and $a_{\text{WC}}$ is the lattice constant, the crystal melts. Following Ref. \cite{ulinich1979phase}, we compute the quantum ($T=0$) uncertainty in the position of an electron in a Wigner crystal subjected to a magnetic field within the harmonic approximation as
\begin{align}\label{eq:posUncertainty}
    \begin{split}
        \langle \bm{u}^2\rangle = \frac{\hbar}{2m}\frac{1}{N}\sum_{\bm{k}} \frac{(\omega_{1,\bm{k}} + \omega_{2,\bm{k}})^2}{\omega_{1,\bm k}\omega_{2,\bm k}}\frac{1}{\sqrt{\omega_c^2 + (\omega_{1,\bm{k}} + \omega_{2,\bm{k}})^2}}.
    \end{split}
\end{align}
Here, $\omega_{1,\bm{k}}$ and $\omega_{2,\bm{k}}$ are the Wigner crystal's longitudinal and transverse phonon frequencies at $B=0$, which we calculate following Ref. \cite{bonsall1977some}. We show the phonon dispersion at both zero and nonzero magnetic field strength in Fig. \ref{fig:lindemannCurve}(a,b). In Fig.~\ref{fig:lindemannCurve}(c), we show the Lindemann melting curve according to Eq.~\ref{eq:posUncertainty}. By inspection, Eq. \ref{eq:posUncertainty} implies that $\langle \bm{u}^2 \rangle$ monotonically decreases as a function of $\omega_c \propto B$ and, therefore, that increasing $B$ never melts a Wigner crystal.

This effect can be understood by approximating each electron as an uncoupled oscillator in a magnetic field and a harmonic potential coming from its interactions with all other electrons. In this approximation, the wavefunction of each electron is a Gaussian $\propto \exp(-\frac{m \sqrt{\omega_0^2 + (\frac{eB}{2mc})^2}}{2\hbar}r^2)$ whose position uncertainty strictly decreases as a function of $B$ ($\omega_0$ is the frequency of the Einstein oscillator at $B=0$).

Ultimately, the Lindemann criterion fails to capture quantum melting of a $B=0$ Wigner crystal into an integer quantum Hall state because it is based on the density distribution of the crystal state rather than the energies of the liquid and crystal states, which ultimately dictate the phase boundary.

\section{Static structure factor}

\subsection{Kohn's theorem and long-wavelength static structure factor}\label{subsec:kohn}

We consider the standard Hamiltonian for a homogeneous 2DEG in a magnetic field
\begin{align}
    H = \sum_i \frac{\bm{\pi}_i^2}{2m} + \sum_{i<j} U(\bm{r}_i - \bm{r}_j),
\end{align}
with $\bm \pi _i = \bm p_i + \frac{e}{c}\bm A (\bm r_i)$ and $U(\bm r)$ a two-body interaction potential. This Hamiltonian decomposes as $H = H_{\rm CM} + H_r$ where
\begin{align}
    H_{\rm CM} = \frac{\bm{\Pi}^2}{2Nm} = \hbar\omega_c\!\left(a_{\rm cm}^\dagger a_{\rm cm}+\frac{1}{2}\right)
\end{align}
with $\bm{\Pi}=\sum_i\bm{\pi}_i$ and $a_{\rm cm}=\frac{\ell}{\hbar\sqrt{2N}}(\Pi_x+i\Pi_y)$ \cite{kohn1961cyclotron}. Thus, the center-of-mass Hamiltonian is that of a charged particle in a magnetic field. The relative-motion Hamiltonian is
\begin{align}
    H_r = \frac{1}{2Nm}\sum_{j<i}(\bm{\pi}_i-\bm{\pi}_j)^2 + \sum_{i<j}U(\bm{r}_i-\bm{r}_j).
\end{align}
Because $[H_{\rm CM},H_r]=0$, the ground state takes the form $|{\rm GS}\rangle = |0_{\rm cm},0_r\rangle$, where $|0_{\rm cm}\rangle$ is the CM ground state and $|0_r\rangle$ is the relative-motion ground state.

We now show that the ground state static structure factor,
\begin{align}
    S(\bm{q}) = \frac{1}{N}\langle{\rm GS}|\rho_{-\bm{q}}\rho_{\bm{q}}|{\rm GS}\rangle,
\end{align}
obeys
\begin{align}\label{eq:kohn}
    \lim_{q\to 0}\frac{S(\bm{q})}{q^2} = \frac{\ell^2}{2}
\end{align}
where $\ell = \sqrt{\frac{\hbar c}{e B}}$ is the magnetic length. For simplicity, we assume full rotational symmetry. Resolving the identity gives
\begin{align}
    S(\bm{q}) = \frac{1}{N}\sum_{n_{\rm cm},n_r}\left|\langle n_{\rm cm},n_r|\rho_{\bm{q}}|0_{\rm cm},0_r\rangle\right|^2.
\end{align}
We combine the continuity equation with the Heisenberg equation of motion to obtain
\begin{align}
    -\bm{q}\cdot\bm{j}_{\bm{q}} = \frac{1}{\hbar}[H,\rho_{\bm{q}}],
\end{align}
where $\rho_{\bm{q}}=\sum_i e^{-i\bm{q}\cdot\bm{r}_i}$ and $\bm{j}_{\bm{q}} = \frac{1}{2m}\sum_i\{\bm{\pi}_i,e^{-i\bm{q}\cdot\bm{r}_i}\}$. This identity allows us to write
\begin{align}
    \langle n_{\rm cm},n_r|\rho_{\bm{q}}|0_{\rm cm},0_r\rangle
    = \frac{-\hbar}{E_{n_{\rm cm},n_r}-E_{0_{\rm cm},0_r}}\langle n_{\rm cm},n_r|\bm{q}\cdot\bm{j}_{\bm{q}}|0_{\rm cm},0_r\rangle.
\end{align}
In the limit $q\to 0$, $j_q^x\to j^x = \frac{1}{m}\Pi^x$. From the definition of $a_{\rm cm}$, we have $\Pi^x = \sqrt{\frac{N}{2}}\frac{\hbar}{\ell}(a_{\rm cm}+a_{\rm cm}^\dagger)$, so
\begin{align}
    \frac{1}{m}\Pi^x = \frac{1}{m}\sqrt{\frac{N}{2}}\frac{\hbar}{\ell}(a_{\rm cm}+a_{\rm cm}^\dagger).
\end{align}
Therefore,
\begin{align}
    \lim_{q\to 0}\frac{1}{q^2}S(q\hat{x})
    &= \frac{1}{N}\sum_{n_{\rm cm},n_r}\left|\frac{-\hbar}{E_{n_{\rm cm},n_r}-E_{0_{\rm cm},0_r}}\left\langle n_{\rm cm},n_r\left|\frac{1}{m}\sqrt{\frac{N}{2}}\frac{\hbar}{\ell}(a_{\rm cm}+a_{\rm cm}^\dagger)\right|0_{\rm cm},0_r\right\rangle\right|^2 \notag\\
    &= \frac{1}{N}\left(\frac{-\hbar}{\hbar\omega_c}\frac{1}{m}\sqrt{\frac{N}{2}}\frac{\hbar}{\ell}\right)^{\!2} \notag\\
    &= \frac{\ell^2}{2},
\end{align}
where only the $n_{\rm cm}=1$, $n_r=0_r$ intermediate state contributes (via $a_{\rm cm}^\dagger|0_{\rm cm}\rangle = |1_{\rm cm}\rangle$), and we have used $\ell^2=\frac{\hbar}{m\omega_c}$.

\subsection{Constraint on long-wavelength Jastrow pseudopotential from Kohn's theorem and the plasma analogy}\label{subsec:plasmaAnalogy}

Here we combine Eq.~\eqref{eq:kohn} with the plasma analogy to show that the Jastrow pseudopotential $u (\bm r)$ must obey
\begin{align}\label{eq:uofr_constraint}
    \lim_{\bm{q}\to 0} q^2 u(\bm{q}) = 0.
\end{align}

We consider the Slater-Jastrow wavefunction
\begin{align}
    \Psi(R) = e^{\sum_{i<j} u(\bm{r}_i - \bm{r}_j)}\,\Psi_{\rm LLL}(R).
\end{align}
Here $\Psi_{\rm LLL}(R)$ is the Slater-determinant wavefunction of the full lowest Landau level. On a disk,
\begin{align}
    \Psi_{\rm LLL}(R) = \prod_{i<j}(z_i - z_j)\,e^{-\frac{1}{4\ell^2}\sum_i \bm r_i^2}.
\end{align}
By the plasma analogy \cite{laughlin1983anomalous,giuliani2008quantum}, the absolute square of this wavefunction maps to the Boltzmann weight of a classical system with potential energy
\begin{align}
    U(R) &\equiv -\frac{1}{\beta}\ln\!\left(|\Psi(R)|^2\right) \notag\\
    &= -\frac{1}{\beta}\!\left(2\sum_{i<j}\!\left[u(\bm{r}_i-\bm{r}_j)+\ln(|\bm{r}_i-\bm{r}_j|)\right]-\frac{1}{2\ell^2}\sum_i \bm r_i^2\right).
\end{align}
Setting $\beta=2$ gives
\begin{align}
    U(R) = \sum_{i<j}\!\left[-u(\bm{r}_i-\bm{r}_j)-\ln(|\bm{r}_i-\bm{r}_j|)\right]+\frac{1}{4\ell^2}\sum_i \bm r_i^2.
\end{align}
$|\Psi(R)|^2$ thus maps to the Boltzmann weight of a classical two-dimensional one-component plasma with a neutralizing background and an additional interparticle potential.
From the classical fluctuation-dissipation theorem, we know
\begin{align}
    S(\bm{q}) = -\frac{1}{\beta n}\chi(\bm{q})
\end{align}
where $\chi(\bm{q})$ is the density-density response function. We can express the density-density response function of the classical plasma in terms of its polarization function $\Pi$ (also known as the proper density-density response function) as
\begin{align}
    \chi(\bm{q}) = \frac{\Pi(\bm{q})}{1-\tilde{u}(\bm{q})\Pi(\bm{q})}
\end{align}
where
\begin{align}
    \tilde{u}(\bm{r}) = -u(\bm{r}) - \ln(|\bm{r}|)
\end{align}
is an effective potential.
The $\bm{q}\to 0$ limit of the polarization function is related to the proper compressibility of the classical system through the compressibility sum rule:
\begin{align}
    \lim_{\bm{q}\to 0}\Pi(\bm{q}) = -n^2 K
\end{align}
where $K=\frac{1}{n^2}\frac{\partial n}{\partial\mu}$ is the proper compressibility. In particular, as long as $\tilde u(\bm q) \Pi (\bm q)$ diverges as $\bm q \rightarrow 0$,
\begin{align}
    \chi(\bm{q}\to 0) = -\frac{1}{\tilde{u}(\bm{q})} + \ldots
\end{align}
and (with $n=\frac{1}{2\pi\ell^2}$)
\begin{align}
    S(\bm{q}\to 0) &= \frac{1}{2n\tilde{u}(\bm{q})} + \ldots \notag\\
    &= \frac{\pi\ell^2}{\tilde{u}(\bm{q})} + \ldots
\end{align}
For the Slater-Jastrow wavefunction to be consistent with Eq.~\eqref{eq:kohn}, the Jastrow factor pseudopotential must obey
\begin{align}
    \lim_{\bm{q}\to 0} q^2\tilde{u}(\bm{q}) = 2\pi.
\end{align}
Given the Fourier transform of the effective potential $\tilde{u}(\bm{q}) = -u(\bm{q}) + \frac{2\pi}{q^2}$, this implies Eq.~\eqref{eq:uofr_constraint}.

\subsection{Classical magnetoplasmons, single mode approximation, and static structure factor}

We consider a 2D fluid of particles with charge $-e$ subject to an out-of-plane magnetic field. It obeys the equation of motion
\begin{align}
    \partial_t\bm{j} = -\frac{ne^2}{m}\bm{\nabla}\phi+\frac{e}{mc}\bm{B}\times\bm{j}
\end{align}
with $\bm{B}=B\hat{z}$. We will work to linear order in $\delta n (\bm r, t)$ where $n(\bm r, t) = n + \delta n (\bm r, t)$. In Fourier space,
\begin{align}
    -i\omega\bm{j}_{\bm{q}} = -i\bm{q}\frac{ne^2}{m}v(q)\rho_{\bm{q}}+\omega_c\hat{z}\times\bm{j}_{\bm{q}}.
\end{align}
Taking another time derivative and applying the continuity equation gives
\begin{align}
    -\omega^2\bm{j}_{\bm{q}} = -\frac{ne^2}{m}v(q)\bm{q}(\bm{q}\cdot\bm{j}_{\bm{q}})-i\omega_c\omega\,\hat{z}\times\bm{j}_{\bm{q}}.
\end{align}
Nontrivial solutions must have frequencies that satisfy
\begin{align}
    \det\!\left[-\omega^2\delta_{ab}+\frac{ne^2}{m}v(q) q_a q_b+i\omega_c\omega\varepsilon_{ab}\right]=0.
\end{align}
The solutions are $\omega=0$ and
\begin{align}\label{eq:classicalmagnetoplasmon}
    \omega_{p}(q) = \sqrt{\omega_c^2+\frac{ne^2}{m}v(q) q^2} = \sqrt{\omega_c^2+\frac{2\pi ne^2}{m}q}
\end{align}
for a Coulomb interaction.

The single-mode approximation gives a rigorous upper bound on the lowest excitation energy $\Delta(q)$ at momentum $q$ in a translation-invariant system \cite{feynman1954atomic}:
\begin{align}
    \Delta(\bm q) \leq \Delta_{\text{SMA}}(\bm q) = \frac{\hbar^2 q^2}{2m S(\bm q)}.
\end{align}
Let us assume that this bound is saturated and that $\Delta(\bm q)=\hbar\omega_p(\bm q)$ at long wavelengths. This assumption implies
\begin{align}
    \begin{split}
        S(\bm q \ll \sqrt n ) &= \frac{\hbar q^2}{2m}\frac{1}{\sqrt{\omega_c^2+\frac{2\pi ne^2}{m}q}} \\ 
        &= \frac{\ell^2 q^2}{2} - \kappa \nu \frac{\ell^3 |\bm q|^3}{4}  + \ldots,
    \end{split}
\end{align}
which is Eq. (8) of the main text. We will now show that the random phase approximation predicts the same small $q$ behavior of the static structure factor.

\subsection{Static structure factor within the random phase approximation}\label{subsec:SRPA}

From the fluctuation-dissipation theorem, the static structure factor is related to the imaginary part of the density-density response function as
\begin{align}
    S(\bm{q}) = -\frac{\hbar}{\pi n}\int_0^\infty d\omega\,\mathrm{Im}\,\chi(\bm{q},\omega).
\end{align}
This motivates the definition of the random phase approximation (RPA) static structure factor \cite{giuliani2008quantum}:
\begin{align}
    S_{\mathrm{RPA}}(\bm{q}) = -\frac{\hbar}{\pi n}\int_0^\infty d\omega\,\mathrm{Im}\,\chi_{\mathrm{RPA}}(\bm{q},\omega).
\end{align}
The density-density response function is related to the polarization function (also known as the proper density-density response function) by
\begin{align}
    \chi(\bm{q},\omega) = \frac{\Pi(\bm{q},\omega)}{1-v(\bm q)\Pi(\bm{q},\omega)}.
\end{align}
In the RPA, the polarization function is set to $\Pi_{\mathrm{RPA}}(\bm{q},\omega) = \chi_0(\bm{q},\omega)$, where
\begin{align}
    \chi_0(\bm{q},\omega) = \frac{1}{A}\sum_{\alpha,\beta}\frac{n_\alpha-n_\beta}{\hbar\omega-(\varepsilon_\beta-\varepsilon_\alpha)+i\hbar\eta}\,|\langle\alpha|\rho(\bm{q})|\beta\rangle|^2
\end{align}
is the density-density response function of the non-interacting system, and $\alpha,\beta$ label single-particle eigenstates. For the Landau-level problem, the single-particle eigenstates are $|n\bm{k}\rangle$, where $n$ is a Landau-level index and $\bm{k}$ is a wavevector in the magnetic Brillouin zone. Their energies are $(n+\frac{1}{2})\hbar\omega_c$, and the squared density-fluctuation matrix elements are \cite{reddy2024non}
\begin{align}
    |\langle m\bm{p}|\rho(\bm{q})|n\bm{k}\rangle|^2 = e^{-\ell^2 q^2/2}|G_{mn}(\bm{q})|^2\,\delta_{\bm{p},[\bm{k}-\bm{q}]}.
\end{align}
Here
\begin{align}
    |G_{mn}(\bm{q})|^2 = \frac{n!}{m!}\!\left(\frac{\ell q}{\sqrt{2}}\right)^{\!2(n-m)}\!\!\left[L_m^{n-m}\!\!\left(\frac{\ell^2 q^2}{2}\right)\right]^2\quad(n\geq m).
\end{align}
If $m>n$, the correct expression is obtained by swapping $n$ and $m$ on the right hand side. For $\nu$ full Landau levels,
\begin{align}
    \chi_0(\bm{q},\omega) = \frac{N_\phi}{\hbar A}\sum_{\substack{m\leq(\nu-1)\\n\geq\nu}} e^{-\ell^2 q^2/2}|G_{nm}(\bm{q})|^2\left[\frac{1}{\omega-(n-m)\omega_c+i\eta}-\frac{1}{\omega-(m-n)\omega_c+i\eta}\right].
\end{align}
Noting that $L_n^m(0)=\binom{n+m}{n}$, we have
\begin{align}
    |G_{mn}(\bm{q})|^2 = \frac{n\ell^2 q^2}{2}\delta_{n,m+1}+\mathcal{O}(q^4).
\end{align}
Therefore,
\begin{align}
    \chi_0(\bm{q}\to 0,\omega) = \frac{1}{2\pi\hbar}\frac{\nu q^2}{2}\left[\frac{1}{\omega-\omega_c+i\eta}-\frac{1}{\omega+\omega_c+i\eta}\right]+\mathcal{O}(q^4)
    \equiv \frac{\nu q^2}{4\pi\hbar}f(\omega)+\mathcal{O}(q^4).
\end{align}
Given $v(q)=\frac{2\pi e^2}{q}$, we have
\begin{align}
    \chi_{\mathrm{RPA}}(\bm{q}\to 0,\omega) &= \chi_0(q,\omega)+v(q)\chi_0^2(q,\omega)+\mathcal{O}(q^4) \notag\\
    &= \frac{\nu q^2}{4\pi\hbar}f(\omega)+\frac{2\pi e^2}{q}\left(\frac{\nu q^2}{4\pi\hbar}f(\omega)\right)^{\!2}+\mathcal{O}(q^4) \notag\\
    &= \frac{\nu q^2}{4\pi\hbar}f(\omega)+\frac{e^2\nu^2 q^3}{8\pi\hbar^2}f^2(\omega)+\mathcal{O}(q^4).
\end{align}
Therefore
\begin{align}
    S_{\mathrm{RPA}}(q) &= -\frac{\hbar}{\pi}\frac{2\pi\ell^2}{\nu}\left[\frac{\nu q^2}{4\pi\hbar}\int_0^\infty d\omega\,\mathrm{Im}\,f(\omega)+\frac{e^2\nu^2 q^3}{8\pi\hbar^2}\int_0^\infty d\omega\,\mathrm{Im}\,f^2(\omega)\right]+\mathcal{O}(q^4) \notag\\
    &= -\left[\frac{\ell^2 q^2}{2\pi}\int_0^\infty d\omega\,\mathrm{Im}\,f(\omega)+\frac{e^2\nu\ell^2 q^3}{4\pi\hbar}\int_0^\infty d\omega\,\mathrm{Im}\,f^2(\omega)\right]+\mathcal{O}(q^4) \notag\\
    &= \frac{\ell^2 q^2}{2}-\frac{e^2\nu\ell^2 q^3}{4\pi\hbar}\int_0^\infty d\omega\,\mathrm{Im}\,f^2(\omega)+\mathcal{O}(q^4) \\
    &= \frac{\ell^2 q^2}{2}-\kappa\nu\frac{\ell^3 q^3}{4}+\mathcal{O}(q^4)
\end{align}
where $\kappa\equiv\frac{e^2}{\hbar\omega_c\ell}$ and we have used $\int_0^\infty \!\!d\omega\;\mathrm{Im}\,f^2(\omega) = \frac{\pi}{\omega_c}$. Analyzing the poles of $\chi_{\mathrm{RPA}}$ with the small-$q$ approximations used above reveals that the magnetoplasmon (also known as magnetoexciton) dispersion at small $q$ obeys
\begin{align}
    \frac{\Delta(q)}{\hbar\omega_c} = 1+\frac{\kappa\nu\ell}{2}q+\ldots
\end{align}
This agrees with the full time-dependent Hartree-Fock analysis of Refs. \cite{kallin1984excitations,macdonald1985hartree} and with the classical magnetoplasmon dispersion Eq. \ref{eq:classicalmagnetoplasmon}. 

\section{Extended methods}

\subsection{Magnetic Bloch orbitals}\label{subsec:magOrbitals}

Here we discuss magnetic Bloch orbitals, which have previously been discussed elsewhere \cite{herzog2022magnetic,wang2021exact,reddy2024non}. These orbitals comprise a complete basis for each Landau level and simultaneously diagonalize magnetic translations on a magnetic lattice enclosing one flux quantum per unit cell.

We consider the Hamiltonian of a single electron with charge $-e$ confined to a two-dimensional plane in a homogeneous, out-of-plane magnetic field,

\begin{align}
    \begin{split}
        H &= \frac{\bm \pi^2}{2m}
    \end{split}
\end{align}
with $\bm \pi = \bm p + \frac{e}{c}\bm A(\bm r)$. We set $\bm{B}(\bm r) = \bm \nabla \times \bm A = -B\hat{z}$. Defining the ladder operators
\begin{align}
    \begin{split}
        a &= \frac{\ell}{\hbar\sqrt{2}}\left(\pi_x + i\pi_y\right) \\
        a^\dagger &= \frac{\ell}{\hbar\sqrt{2}}\left(\pi_x - i\pi_y\right),
    \end{split}
\end{align}
which obey the algebra
\begin{align}
    \begin{split}
        [a,a^\dagger]=1,
    \end{split}
\end{align}
we have
\begin{align}
    \begin{split}
        H = \hbar\omega_c\left(a^\dagger a+\frac{1}{2}\right).
    \end{split}
\end{align}
The condition $2\pi \ell^2 B = \Phi_0 = \frac{hc}{e}$ defines the magnetic length $\ell$. The magnetic translation operator is
\begin{align}
    \begin{split}
        t(\bm{d}) = e^{i\chi_{\bm d} (\bm r)}T(\bm d),
    \end{split}
\end{align}
where
$\bm \nabla \chi_{\bm d}(\bm r) = \frac{2\pi}{\Phi_0}\left[\bm{A}(\bm r + \bm d) - \bm A (\bm r)\right]$. $T(\bm d) = e^{i\bm{p\cdot\bm{d}}/\hbar }$ is an ordinary translation operator. Magnetic translations obey
\begin{align}\label{eq:commtH}
    [t(\bm d),\bm\pi]=0= [t(\bm d),H]
\end{align}
and
\begin{align}\label{eq:commtt}
    \begin{split}
        t(\bm d')t(\bm d) = e^{i\bm{d}'\wedge\bm d/(2 \ell^2)}t(\bm d' + \bm d) = e^{i\bm{d}'\wedge\bm d/ \ell^2}t(\bm d)t(\bm d').
    \end{split}
\end{align}
We also define the guiding center operator $\bm R = \bm r + \frac{\ell^2}{\hbar} \hat z \times \bm \pi$, which obeys $[R_a,\pi_b]=0$.

We seek an eigenbasis of $H$ obeying the boundary condition
\begin{align}\label{eq:SCBC}
    \begin{split}
        t(\bm L_i)\ket{\phi} = \ket{\phi}
    \end{split}
\end{align}
where $\bm L = l_1 \bm{L}_1+l_2\bm{L}_2$ is a supercell lattice vector. $\bm{L}_{1,2}$ are primitive vectors of the supercell lattice. Let $N_{\Phi} = |\bm L_1 \times \bm L_2|/(2\pi\ell^2)$ and $N_1$, $N_2$ be integers such that $N_1 N_2=N_{\Phi}$. Let $\bm{a}_i=\bm{L}_i/N_{i}$ so that $|\bm{a}_1\times \bm{a}_2| = 2\pi\ell^2$, implying $[t(\bm a_1),t(\bm a_2)]=0$. $\bm{a}_{1,2}$ are primitive vectors of the magnetic lattice. We define the state $\ket{f_0}$ by the conditions $a\ket{f_0}=0$ and $(R_x-iR_y)\ket{f_0}=0$. We also define $\ket{f_n} = \frac{1}{\sqrt{n!}}(a^{\dagger})^n \ket{f_0}$.

We seek to construct magnetic Bloch states $\ket{\phi_{n \bm k}}$ that obey the eigenvalue equations
\begin{align}\label{eq:MagBlochBC}
    \begin{split}
        t(\bm a_i) \ket{\phi_{n\bm{k}}} = e^{i\bm k \cdot \bm a_i}\ket{\phi_{n \bm k}}
    \end{split}
\end{align}
and
\begin{align}\label{eq:HamEval}
    \begin{split}
        a^\dagger a\ket{\phi_{n \bm k}} = n\ket{\phi_{n\bm k}}.
    \end{split}
\end{align}
To proceed, we first define
\begin{align}
    \begin{split}
     \ket{\phi_{n\bm 0}} &= \sum_{r,s\in \mathbb{Z}}t(r \bm a_1)t(s \bm{a}_2)\ket{f_n} \\
     &= \sum_{r,s\in \mathbb{Z}}e^{i\pi rs}t(r \bm a_1+s \bm{a}_2)\ket{f_n}
    \end{split}
\end{align}
and then define
\begin{align}
    \begin{split}
        \ket{\phi_{n\bm k}} &= t(\ell^2 \hat{z} \times \bm k)\ket{\phi_{n\bm 0}} \\
        &= \sum_{r,s\in \mathbb{Z}}e^{-i\bm k \cdot (r\bm a_1 +s\bm{a}_2)}e^{i\pi rs}t(r \bm a_1+s \bm{a}_2) t(\ell^2 \hat{z}\times \bm k)\ket{f_n}
    \end{split}
\end{align}
Given these definitions, Eq. \ref{eq:commtH} implies Eq. \ref{eq:HamEval} and Eq. \ref{eq:commtt} implies Eq. \ref{eq:MagBlochBC}.

 To make Eqs.~\ref{eq:SCBC} and \ref{eq:MagBlochBC} compatible, $\bm k$ can take any value $\bm {k} = k_1 \bm{G}_1 + k_{2}\bm{G}_2$ where $\bm{L}_i\cdot\bm{G}_j=2\pi\delta_{ij}$. There are $N_{\Phi}$ unique values of $\bm{k}$, with integral $0\leq k_i < N_i$, in the first magnetic Brillouin zone. The magnetic Bloch states thus comprise a complete basis of $N_{ \Phi}$ states per Landau level.

For numerical calculations, we use the symmetric gauge $\bm A (\bm r) = \frac{B}{2}(+y,-x)$. In this gauge, $\chi_{\bm d}(\bm r) = \bm{r}\wedge \bm{d}/(2\ell^2)$,
\begin{align}\label{eq:phi_n}
    \begin{split}
        f_n(\bm r) = \frac{1}{\ell\sqrt{2n!}}(i\bar{z})^ne^{-\bm r^2/(4\ell^2)}
    \end{split}
\end{align}
and
\begin{align}
    \begin{split}
        \phi_{n\bm k}(\bm r) = \sum_{\bm{a}} e^{i(\frac{\bm k \cdot (\bm r -\bm a)}{2} + \frac{\bm r \wedge \bm a}{2\ell^2}  + \pi rs )}f_{n}(\bm r + \bm{a}+ \ell^2\hat{z}\times \bm{k})
    \end{split}
\end{align}
where $\bm a = r\bm a_1 + s\bm a_2$. This infinite sum over magnetic lattice vectors $\bm{a}$ converges rapidly due to the Gaussian factor in Eq. \ref{eq:phi_n}.

The identity
\begin{align}\label{eq:magTransRel}
    \begin{split}
        \phi_{n\bm{k}}(\bm r + \bm a) &= e^{i(\bm k \cdot (\bm a + \bm r/2) - \frac{\bm r \wedge \bm a}{2 \ell^2} -\pi rs)}\phi_{n\bm{0}}(\bm r - \ell^2 \bm{k}\times \hat{z})
    \end{split}
\end{align}
is useful for numerical calculation because it allows one to easily calculate the value of $\phi_{n \bm k}$ at any position knowing only $\phi_{n\bm{0}}$ in a single magnetic unit cell. Exploiting this formula, we precompute $\phi_{n\bm 0}$ on a fine grid ($100\times 100$ points) in a single magnetic unit cell and interpolate its real and imaginary parts with two-dimensional cubic $B$-spline interpolants on this grid \cite{Kittisopikul_Interpolations_jl_2025}. We then combine this interpolation with Eq. \ref{eq:magTransRel} to quickly compute $\phi_{n \bm k}(\bm r)$.

We note that the algebraic relationship $\pi_x \ket{\phi_{n\bm{k}}} = \frac{\hbar}{\ell\sqrt{2}}(a^{\dagger}+a)\ket{\phi_{n\bm{k}}} = \frac{\hbar}{\ell\sqrt{2}}\left[\sqrt{n+1}\ket{\phi_{(n+1)\bm{k}}}+\sqrt{n}\ket{\phi_{(n-1)\bm{k}}}\right]$ (and similar for $\pi_y$) is useful for computing the local energy.

\subsection{Plane wave expansion at $B=0$}\label{subsec:planewave}

At $B=0$, we impose periodic boundary conditions and use a plane wave basis for the single-particle orbitals:
\begin{align}
    \begin{split}
        \ket{\psi_{n\bm k}}=\sum_{m}U^{\bm k}_{nm}\ket{\bm k + \bm g_m}
    \end{split}
\end{align}
where $\bra{\bm r}\ket{\bm k}= e^{i\bm{k}\cdot\bm{r}}$. For the Fermi liquid ansatz, each electron occupies one of the $N$ lowest-energy plane waves and there is no variational freedom in the orbitals. For $N=121$, this set is unique, i.e. the system is ``closed shell''. For the Wigner crystal, we choose a Wigner-Seitz Brillouin zone mesh. For each point on this mesh, we expand a Bloch orbital in a plane wave basis out to the second shell of reciprocal lattice vectors with 13 reciprocal lattice vectors in total (see Fig. \ref{fig:meshDiagrams}).

\subsection{Supercell geometry}\label{subsec:supercell}

\begin{figure}
    \centering\includegraphics[width=0.8\linewidth]{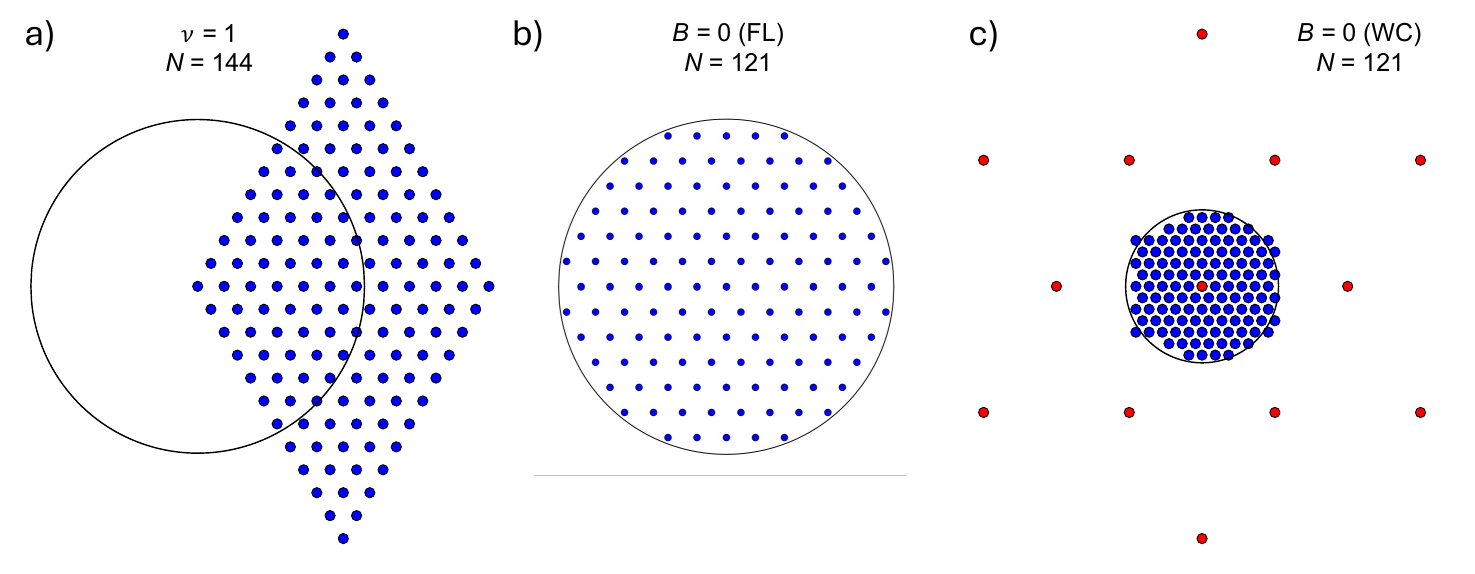}
    \caption{Diagrams of momentum-space meshes used in VMC calculations. The black circle is centered about the origin and has radius $k_F = \sqrt{4\pi n}$. The different panels are not to scale. Blue dots denote the mesh of $\bm k$ points and the red dots denote reciprocal lattice vectors $\bm{g}$, which are only relevant to the $B=0$ Wigner crystal.}
    \label{fig:meshDiagrams}
\end{figure}

We use a hexagonal magnetic unit cell with $\bm{a}_1 = \nu a_{\text{WC}}(\frac{\sqrt{3}}{2},-\frac{1}{2})$ and $\bm{a}_2 = a_{\text{WC}}(\frac{\sqrt{3}}{2},+\frac{1}{2})$ where $a_{\text{WC}}=\sqrt{\frac{2}{\sqrt{3}n}}$. $\nu$ appears here so that the magnetic unit cell has an aspect ratio of $\nu$ and is commensurate with the Wigner crystal lattice. For a system with $N$ electrons, we set $N_1 = \sqrt{N}/\nu$ and $N_2=\sqrt{N}$ so that the supercell always has an equal aspect ratio. An example of a magnetic Brillouin zone mesh is shown in Fig. \ref{fig:meshDiagrams}(a).

At $B=0$, the same setup applies upon replacing $\nu\rightarrow 1$. Momentum space meshes for the $B=0$ Fermi liquid and Wigner crystal ansätze are shown in Fig. \ref{fig:meshDiagrams}(a,b). We remark that, at $B\neq 0$, all valid magnetic Brillouin zones give the same Hilbert space. In contrast, at $B=0$, the choice of mesh clearly matters for the Fermi liquid phase and also matters for the Wigner crystal phase because of the finite reciprocal lattice truncation.

\subsection{Gaussian orbitals}\label{subsec:gaussian}

At $B=0$, we define Gaussian orbitals obeying periodic boundary conditions as
\begin{align}
    \psi_{\bm R}(\bm r) &= \frac{1}{\sqrt{2\pi}\ell_0}\sum_{\bm L}e^{-\frac{(\bm r - \bm R + \bm L)^2}{4 \ell_0^2}}
\end{align}
where $\bm R$ is the center of the Gaussian within the supercell, $\bm L$ are supercell lattice vectors, and $\ell_0$ is a variational parameter.

At $B\neq 0$, we define Gaussian orbitals obeying magnetic boundary conditions as
\begin{align}
    \begin{split}
        \ket{\psi_{\bm R}} &= \sum_{n,m \in \mathbb{Z}} t(n\bm L_1)t(m\bm L_2) \ket{\phi} \\
        &= \sum_{\bm L} e^{i\pi N_{\Phi} nm} t(\bm L) \ket{\phi}
    \end{split}
\end{align}
where $\bm L = n\bm L_1 + m \bm L_2$. Here,
\begin{align}
    \phi(\bm r) = \braket{\bm r}{\phi} = \frac{1}{\sqrt{2\pi} \ell_0 }e^{-\frac{\bm r^2}{4\ell_0^2}}.
\end{align}
In the symmetric gauge,
\begin{align}
    t(\bm d) = e^{i\frac{\bm r \wedge \bm d}{2\ell^2}}T(\bm d)
\end{align}
and
\begin{align}
    \begin{split}
        \psi_{\bm R} (\bm r) &= \frac{1}{\sqrt{2\pi} \ell_0 }\sum_{\bm L} e^{i\left(\pi N_{\Phi}nm  + \frac{\bm r \wedge \bm L}{2\ell ^2} \right )}  e^{-\frac{(\bm r - \bm R + \bm L)^2}{4\ell_0^2}}.
    \end{split}
\end{align}

\subsection{Jastrow factor}
Our two-body Jastrow factor takes the form
\begin{align}
    \begin{split}
        J(R) = \exp\left(\sum_{i<j}u(\bm r_i-\bm r_j)\right)
    \end{split}
\end{align}
where
\begin{align}
    \begin{split}
        u(\bm r) &= S(f(\bm r)),
    \end{split}
\end{align}
\begin{align}
    \begin{split}
        f(\bm r) &= \frac{2}{3}\sqrt{\sum_{i=1,2,3}\sin^2(\bm G_i \cdot \bm r/2)},
    \end{split}
\end{align}
and
\begin{align}
    \begin{split}
        S(x) &= \sum_{m=0}^{M}c_mB_3(Mx-m).
    \end{split}
\end{align}
Here
\begin{equation}
  B_3(x) =
  \begin{cases}
    \frac{1}{6}\left( 4 - 6 |x+1|^2 + 3|x+1|^3\right) & 0 \leq |x+1| \leq 1, \\
    \frac{1}{6} (2- |x+1|)^3 & 1 \leq |x+1| \leq 2,  \\
    0 & \text{otherwise}
  \end{cases}
\end{equation}
is the cardinal cubic $B$-spline function, which is twice differentiable everywhere and nonzero only in the region $-3\leq x\leq 1$. $f(\bm r)$ is scaled so that $\max f(\bm{r}) = 1$ and is a ``periodification'' of $|\bm r|$ in the sense that (1) $f(\bm r) \propto |\bm{r}|$ at small $|\bm{r}|$ and (2) $f(\bm r + \bm L) = f(\bm r)$ where $\bm L$ is a supercell lattice vector. Here $\bm{G}_1=\frac{2\pi \bm L_2 \times \hat z}{|\bm L_1 \times \bm L_2|}$, $\bm{G}_2=\frac{2\pi \hat{z}\times \bm L_1}{|\bm L_1 \times \bm L_2|}$, and $\bm G_3 = \frac{\bm G_1 + \bm G_2}{2}$ are reciprocal lattice vectors of the supercell. $f(\bm{r})$ is approximately isotropic at small $|\bm r|$ for supercells with an equal aspect ratio, which we always use in this work.

Our parameterization of $u(\bm r)$ is intended to capture both short and long-range correlations in a unified form. It is inspired by the two-body spline Jastrow in QMCPack \cite{kim2018qmcpack} and the Jastrow factor described in the Supplementary Material of Ref. \cite{valenti2024nematic}. The spline function $S(x)$ is highly expressive.

The cusp condition (in this case, for parallel spins in two dimensions)
\begin{align}
    \begin{split}
        \hat{r} \cdot \bm \nabla u(\bm r) &= \frac{1}{3a_B}
    \end{split}
\end{align}
eliminates divergence in the local energy where two particles coincide \cite{kato1957eigenfunctions,foulkes2001quantum}. It imposes the constraint $c_0 = c_2 - \frac{1}{\pi\sqrt{2} M}\frac{|\bm{L}_1|}{a_B}$. $c_1$ through $c_M$ are optimizable parameters and $c_0$ is fixed by the cusp condition. The number $M$ of ``control points'' is a free hyperparameter that we set between 8 and 15.

\subsection{Local energy}

The local energy is defined as
\begin{align}
    \begin{split}
        E_L(R) = \frac{\bra{R} H \ket{\Psi}}{\langle R|\Psi \rangle} = \frac{1}{\Psi(R)} H \Psi(R).
    \end{split}
\end{align}
The kinetic contribution to the local energy of our Slater-Jastrow wavefunction,
\begin{align}
    \begin{split}
        \Psi(R) = \det\left[\psi_i(\bm{r}_j)\right] \exp\left(\sum_{i<j}u(\bm{r}_i-\bm{r}_j)\right) \equiv  D(R) e^{U(R)},
    \end{split}
\end{align}
is
\begin{align}
    \begin{split}
        T_L &= \frac{1}{2m}\sum_i\frac{\bm {\pi}_i^2 \Psi }{\Psi} \\
        &= \frac{1}{2m}\sum_i \left[ \frac{\bm{\pi}_i^2 D}{D} - \hbar^2\nabla_i^2 U - (\hbar \bm \nabla_i U)^2 -2i \hbar \bm \nabla_i U\cdot \frac{\bm \pi_i D}{D}\right]
    \end{split}
\end{align}
where we recall that $\bm \pi = -i\hbar\nabla + \frac{e}{c}\bm A(\bm r)$ and we have suppressed explicit $R$-dependence.
We compute the Coulomb potential contribution using the Ewald method (see, for example, Ref. \cite{geier2025self}).

\subsection{Orbital parameterization}

Our approach to orbital parameterization is inspired by that of Ref. \cite{toulouse2007optimization}. We parameterize our single-particle orbitals as
\begin{align}\label{eq:orbitalRotation}
\begin{split}
\ket{\psi_{\bm{k}m}} = \sum_{n=0}^{n_{\text{max}}} U^{\bm{k}}_{m n}\ket{\phi_{\bm{k}n}}
\end{split}
\end{align}
where
\begin{align}
    \begin{split}
        U^{\bm{k}}= \exp\left(i\Theta^{\bm{k}}\right)
    \end{split}
\end{align}
is the unitary orbital rotation matrix and $\Theta^{\bm{k}}_{mn}$ is a Hermitian matrix. At filling factor $\nu\in \mathbb{Z}$, $\nu$ orbitals are occupied and $(n_{\text{max}}+1-\nu)$ are unoccupied at each $\bm{k}$. The matrix elements $\Theta^{\bm{k}}_{mn}$, where $m$ and $n$ are both in the occupied sector ($< \nu$) or both in the unoccupied sector ($\geq \nu$), generate unitary transformations within the occupied or unoccupied sector whose only effect is to multiply the many-body wavefunction by an overall phase. Therefore, we set them to $0$. We are left with $(\# \text{ of } \bm{k} \text{ points})\times \nu \times (n_{\text{max}}+1-\nu)\times 2$ orbital rotation parameters: $\theta^{\bm k,R}_{mn}$ and $\theta^{\bm k,I}_{mn}$ for $0\leq m <\nu$ and $\nu \leq n \leq n_{\text{max}}$ where
\begin{align}
    \begin{split}
        \Theta^{\bm k}_{mn} = \begin{cases}
            \theta^{\bm k ,R}_{mn}+i\theta^{\bm k, I}_{mn} & \text{if $m<n$} \\
            \theta^{\bm k,R}_{nm}-i\theta^{\bm k,I}_{nm} & \text{if $n<m$}.
        \end{cases}
    \end{split}
\end{align}

To run stochastic reconfiguration (see Ref. \cite{becca2017quantum}), we need the quantity $\partial_{\alpha} \ln \Psi $ where $\alpha = \theta_{nm}^{\bm k, R/I}$ is an orbital parameter. We use the formula
\begin{align}
    \begin{split}
        \frac{\partial \ln \Psi}{\partial \Theta^{\bm k}_{nm}} &= \frac{\partial \ln D}{\partial \Theta^{\bm k}_{nm}} \\
        &= \sum_{p,q}\frac{\partial \ln D}{\partial U^{\bm{k}}_{pq}} \frac{\partial U^{\bm k}_{pq}}{\partial \Theta^{\bm k}_{nm}}
    \end{split}
\end{align}
and we recall that $D = \det \psi_i (\bm r_j)$ is the Slater determinant. Now
\begin{align}
    \begin{split}
        \frac{\partial \ln D}{\partial U^{\bm{k}}_{pq}} &= \frac{\det \tilde \psi_i(\bm r_j)}{D}
    \end{split}
\end{align}
where
\begin{align}
    \begin{split}
        \tilde \psi_i = \begin{cases}
            \phi_{\bm k q} & \text{if  $i = \bm kp$} \\
            \psi_i & \text{else}.
        \end{cases}
    \end{split}
\end{align}
Here $i\sim \bm k n$ is a compound orbital index and $\phi_{\bm k q}$ is a basis orbital (see Eq. \ref{eq:orbitalRotation}). Notice that $\tilde \psi_i (\bm r_j)$ differs from $\psi_i (\bm r_j)$ only in one row, so we can use the Sherman-Morrison fast update formula to compute the ratio of determinants.

From here on we will drop the explicit $\bm k$ index because all terms will share it. From Theorem 4.5 of Ref. \cite{najfeld1995derivatives}, it follows that
\begin{align}
    \begin{split}
        \frac{\partial U_{mn}}{\partial \Theta_{pq}} = \frac{\partial \exp(i\Theta)_{mn}}{\partial \Theta_{pq}} = i\sum_{r,s} V_{mr}V^{\dagger}_{rp}V_{qs}\Phi_{rs}V^{\dagger}_{sn}
    \end{split}
\end{align}
where $V \Lambda V^{\dagger}=i\Theta$ is the spectral decomposition of $i\Theta$ and
\begin{align}
    \begin{split}
        \Phi_{pq} = \begin{cases}
            \frac{e^{\Lambda_{pp}}-e^{\Lambda_{qq}}}{\Lambda_{pp}-\Lambda_{qq}} & \text{if $\Lambda_{pp} \neq \Lambda_{qq}$} \\
            e^{\Lambda_{pp}} & \text{if $\Lambda_{pp} = \Lambda_{qq}$}.
        \end{cases}
    \end{split}
\end{align}
Finally, these derivatives are related to derivatives with respect to the variational parameters as
\begin{align}
    \begin{split}
        \frac{\partial}{\partial \theta^{R}_{nm}} = \frac{\partial}{\partial \Theta_{nm}} + \frac{\partial}{\partial \Theta_{mn}}
    \end{split}
\end{align}
and
\begin{align}
    \begin{split}
        \frac{\partial}{\partial \theta^{I}_{nm}} = i\left (\frac{\partial}{\partial \Theta_{nm}} - \frac{\partial}{\partial \Theta_{mn}} \right).
    \end{split}
\end{align}
These are all the formulae we need to implement our orbital parameterization and its optimization via stochastic reconfiguration.

\subsection{Sampling}

We sample the probability distribution $|\Psi(R)|^2$ using the standard Metropolis-Hastings algorithm, proposing single-particle moves from a Gaussian distribution whose radius is updated between optimization steps to maintain a 50\% acceptance rate \cite{metropolis1953equation}. We use the method of Ref. \cite{ceperley1977monte} to efficiently compute ratios of determinants of Slater matrices that differ only in one row.

During wavefunction optimization, we typically take $\approx 10^4$ samples per optimization step. At later stages of optimization, we take up to $10^6$ samples per optimization step. We sample the wavefunction after every single-particle move. Once optimized, we typically take $10^7$ samples to estimate energies and correlation functions. The autocorrelation time we find for the local energy of our optimized wavefunction is typically a few hundred to a few thousand single-particle moves.

\bibliography{ref}